\documentclass[12pt, draftclsnofoot, onecolumn]{IEEEtran}
\usepackage{stfloats}
\ifCLASSINFOpdf
\else
\fi

\usepackage{CJK}
\usepackage{graphicx}
\usepackage{caption}
\usepackage{epstopdf}
\usepackage{subfigure}
\usepackage[cmex10]{amsmath}
\usepackage[amsmath,thmmarks]{ntheorem}
\usepackage{amssymb}
\usepackage{ntheorem}
\usepackage{mathabx}
\usepackage{algorithm}
\usepackage{algorithmic}
\usepackage{multirow}
\usepackage{xcolor}
\usepackage{multirow}
\usepackage{bm}
\usepackage{mathabx}
\usepackage{stmaryrd}
\usepackage{upgreek}
\newtheorem{theorem}{Theorem}
\newtheorem{corollary}{Corollary}
\begin{document}
\title{On the Degrees of Freedom of MIMO X Networks with Non-Cooperation Transmitters}

\author{Tengda~Ying,~
        Wenjiang~Feng,~Weifeng~Su,~and~Weiheng~Jiang       
\thanks{T. Ying, W. Feng, and W. Jiang are with the College of Communication Engineering, University of Chongqing, Chongqing, 400044, P.R. China (e-mail: tengdaying@cqu.edu.cn; fengwj@cqu.edu.cn; whjiang@cqu.edu.cn).}
\thanks{W. Su is with the Department of Electrical Engineering, State University of New York, Buffalo, NY 14260 USA (e-mail: weifeng@buffalo.edu).}}

\maketitle

\begin{abstract}
Due to limited backhaul/feedback link capacity and channel state information (CSI) feedback delay, obtaining global and instantaneous channel state information at the transmitter (CSIT) is a main obstacle in practice. In this paper, novel transmission schemes are proposed for a class of interference networks that can achieve new trade-off regions between the sum of degrees of freedom (sum-DoF) and CSI feedback delay with distributed and temperately-delayed CSIT. More specifically, a distributed space-time interference alignment (STIA) scheme is proposed for the two-user multiple-input multiple-output (MIMO) X channel via a novel precoding method called \textit{Cyclic Zero-padding}. The achieved sum-DoFs herein for certain antenna configurations are greater than the best known sum-DoFs in literature with delayed CSIT. Furthermore, we propose a distributed retrospective interference alignment (RIA) scheme that achieves more than 1 sum-DoF for the \textit{K}-user single-input single-output (SISO) X network. Finally, we extend the distributed STIA to the \textit{M}$\times$\textit{N} user multiple-input single-output (MISO) X network where each transmitter has \textit{N}$-$1 antennas and each receiver has a single antenna, yielding the same sum-DoF as that in the global and instantaneous CSIT case. The discussion and the result of the MISO X network can be extended to the MIMO case due to spatial scale invariance property.
\end{abstract}


\IEEEpeerreviewmaketitle

\section{Introduction}
\IEEEPARstart{C}{hannel} state information at the transmitter (CSIT) is of great importance in interference alignment in wireless communication. While CSIT can be used to align the interference from multiple transmitters to reduce the aggregate interference footprint in interference networks, the caveat behind most of these results has been the assumption of perfect, sometimes global and instantaneous CSIT \cite{Maddah2008Communication,Jafar2008Degrees,Cadambe2009Interference,Sun2012Degrees,Agustin2012Degrees,Abdoli2012Full,
Wang2011Subspace,Cadambe2008Interference}. Nevertheless, it is difficult to achieve the theoretical gains of these techniques in practice due to the distributed nature of the users and the increasing mobility of wireless nodes. 

It is not practical to obtain instantaneous CSIT when the channel coherence time is shorter than the feedback delay, i.e., completely-delayed CSIT. Thus, the completely-delayed CSIT didn't attract much attention to improve the sum-DoF until Maddah-Ali \textit{et al}. introduced the idea of retrospective interference alignment (RIA), where the receivers can successfully decode appreciable symbols based on the centralized transmitter's ability to reconstruct all the interference seen in previous symbols \cite{Maddah2012Completely}. Extensive works on RIA over interference networks \cite{Maleki2012Retrospective,A2011On,Abdoli2011On,Ghasemi2012On,Tandon2012On} have been carried out following the seminar work \cite{Maddah2012Completely}, especially the recent works on the interference alignment with delayed CSIT \cite{Vaze2012The,Yang2012Degrees,Lashgari2013Linear,Kao2014Linear}. The basic approach of dealing with CSI feedback delay for RIA is to seek the possibility of aligning inter-user interference between the past and the currently observed signals by creating new channel side information, with the help of global delayed CSIT. Although there are plenty of works considering various channel models with no CSIT assumption \cite{Vaze2009The,Huang2012On,Jafar2012Blind}, it is interesting that Maleki \textit{et al}. introduced a distributed version of RIA over the interference network, where the three-user interference channel and two-user X-channel with delayed CSIT can respectively achieve more than 1 sum-DoF almost surely \cite{Maleki2012Retrospective}. However, there is still a gap between the achievable bounds of the sum-DoF and the outer bounds that were developed under the full CSIT assumption (perfect, global and instantaneous CSIT). The delayed CSI feedback setting has been naturally extended to some other forms such as delayed output feedback \cite{Tandon2011Degrees} and delayed Shannon feedback \cite{Vaze2011The}, aiming to improve the performance of network. 

On the other hand, obtaining global CSIT is another bottleneck to realize transmitter cooperation with CSIT sharing among distributed transmitters, especially non-collocated transmitters with limited feedback link capacity. Recently, Lee \textit{et al}. introduced the temperately-delayed CSIT regime, which assumes independent and identically distributed (i.i.d.) block fading channels with perfect knowledge of both current and delayed CSIT alternatively \cite{lee2014space}. The concept of the temperately-delayed CSIT has attracted attention due to the reduced CSI feedback and the consideration of the distributed nature of transmitters. In particular, when the transmitters are distributed, each transmitter may obtain local CSI of the channel to its associated receivers using feedback links without exchanging CSI among the transmitters. In the context of the ${K}\times{2}$  X channel with a single antenna at each node, an interesting result was reported in \cite{Lee2014Distributed} that not only the temperately-delayed CSIT has a significant impact on increasing the DoF, but also with \textit{local} delayed CSIT. Interestingly, the result does not hold in the multiuser multiple-input multiple-output (MIMO) interference network \cite{Lee2012CSI}. The interference alignment with local CSIT becomes intriguing and the corresponding maximum achievable degrees of freedom is still unknown in this scenario.

Therefore, a natural question being raised is: Does local CSIT improve DoF in MIMO X wireless networks with non-cooperation distributed transmitters? We would like to address the fundamental problem in this paper. In the context of interference networks, the problem may be addressed by answering the following two questions:  1) Is RIA still able to obtain DoF benefits in interference networks with local CSIT? 2) Is space-time interference alignment (STIA) still able to obtain DoF benefits in multiuser MIMO interference networks with local CSIT? Specifically, we focus on the \textit{M}$\times$\textit{N}  user MIMO X network which has received significant attention in recent years \cite{Maddah2008Communication,Jafar2008Degrees,Cadambe2009Interference,Sun2012Degrees,Agustin2012Degrees,A2011On
,Abdoli2011On,Ghasemi2012On,Tandon2012On,Lee2014Distributed}. Under the full CSIT assumption, the capacity region of the \textit{M}$\times$\textit{N}  user single-input single-output (SISO) X network was characterized in \cite{Cadambe2009Interference}. The sum-DoF of the \textit{M}$\times$\textit{N} user MIMO X network with $A$ antennas at each node was shown to be $\frac{AMN}{M+N-1}$  with full CSIT \cite{Sun2012Degrees}. Also, it was proved that the  \textit{M}$\times$\textit{N} user multiple-input single-output (MISO) X network with $R$ antennas at each transmitter and a single antenna at each receiver almost surely has a sum-DoF of min($N$,$\frac{MNR}{N+MR-R}$). Since then, the impact of delayed CSIT has been actively studied, especially for some SISO and MIMO X channels \cite{A2011On,Abdoli2011On,Ghasemi2012On,Tandon2012On}. In particular, enhanced DoF was achieved for the two-user MIMO X channel by Maleki in \cite{Maleki2012Retrospective} and it was further improved by Ghasemi \textit{et al} in \cite{A2011On}. The DoF of the two-user MIMO X-channel with delayed CSIT was investigated for the symmetric case in \cite{Ghasemi2012On}. Soon thereafter, Abdoli \textit{et al}. investigated the sum-DoF of the ${2}\times{K}$  SISO X channel with delayed CSIT, which converges to the value of $\frac{1}{\ln2}$  as $K$ goes to infinity \cite{Abdoli2011On}. In \cite{Lee2014Distributed}, it was shown that it is possible to strictly increase the sum-DoF with temperately-delayed and local CSIT for the two-user SISO X channel.

In this work, we are able to answer the fundamental questions raised above by characterizing the sum-DoF for \textit{M}$\times$\textit{N} user MIMO X networks with distributed and temperately-delayed CSIT. We first address the scenario of transmission over the two-user MIMO X channel with temperately-delayed local CSIT, where each transmitter has $A$ antennas and each receiver has $B$ antennas. By developing a novel precoding technique, namely \textit{Cyclic Zero-padding}, we obtain new achievable DoFs for the two-user MIMO X channel which is given by $4A/(2+\lceil{\frac{2A-B}{B}}\rceil)$ and it is greater than the best known DoFs in literature. Then, we consider the $K$-user ($M\!\!=\!\!N\!\!=\!\!K$) SISO X network with temperately-delayed and local CSIT. The achievable sum-DoF in this case is $\frac{2(2K-1)}{3K-1}$ which is the same as that in \cite{A2011On}. However, we achieve the sum-DoF via an alternative multiphase transmission scheme which shows that the RIA is still able to obtain the DoF benefits in the interference network with local CSIT. It implies that the local and temperately-delayed CSIT is definitely beneficial to obtain larger sum-DoF compared to the case without CSIT in which the sum-DoF reduces to 1 for the $K$-user SISO X channel. Finally, we extend our results to \textit{M}$\times$\textit{N} user MIMO X network from the perspective of spatial scale invariance by exploring a general setting of MISO X network where each transmitter has $N-1$  antennas and each receiver has a single antenna. An interesting result is that the achievable sum-DoF in this case is $\frac{MN(N-1)}{M(N-1)+1}$ with local CSIT which reaches the outer bound of the full CSIT case. 

The rest of this paper is structured as follows. The system model is described in Section II. Section III presents our main results on the sum-DoF trade-off regions, and the achievable sum-DoFs are compared with those obtained with delayed CSIT, with full CSIT, and without CSIT, respectively. In Sections IV, V and VI, our transmission schemes for the two-user MIMO X channel, $K$-user SISO X network, and \textit{M}$\times$\textit{N}  user MISO X network with local and temperately-delayed CSIT are specified, respectively. Finally, Section VII concludes this paper.

Throughout the paper, we use the following notations. Matrix transpose, inverse and determinant are denoted by \textbf{A}$^T$, \textbf{A}$^{-1}$, and det(\textbf{A}), respectively. We use lowercase letters for scalars, lowercase bold letters for vectors, and uppercase bold letters for matrices. 

\begin{figure}[]
\centering
\includegraphics[width=3.5in]{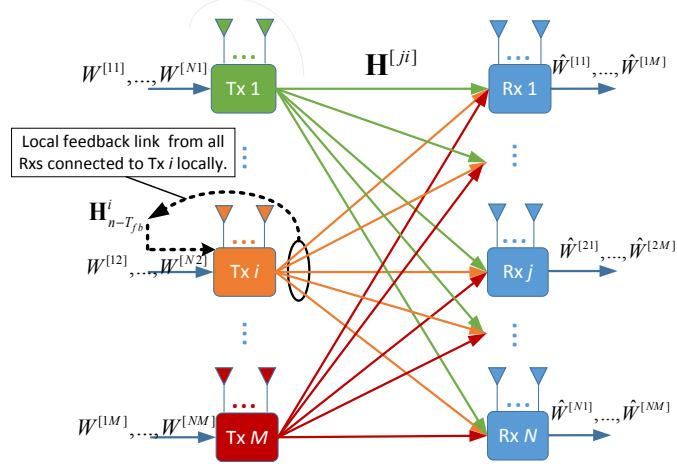}
\caption{Illustration of \textit{M}$\times$\textit{N} user MIMO X network.}
\label{fig.1}
\end{figure}

\section{System Model}

\subsection{Signal Model}
As illustrated in Fig. 1, an \textit{M}$\times$\textit{N} user MIMO X network is a single-hop communication network with $M$ transmitters and $N$ receivers where transmitter $i$ has an independent message $W^{[ji]}$  for receiver $j$, for each $i\in\{1,2,...,M\}$, $j\in\{1,2,...,N\}$. Transmitter $i$ has $A_i$  antennas and receiver $j$ has  $B_j$ antennas. The ${M}\times{N}$  user MIMO X network is described as
\begin{equation}
{{\bf{Y}}^{[j]}}(n) = \sum\limits_{i=1}^{M} {{{\bf{H}}^{[ji]}}(n){{\bf{X}}^{[i]}}(n)}  + {{\bf{Z}}^{[j]}}(n),\;j \in \{ 1,2,...,N\}, 
\end{equation}
where $n$ represents the time slot, ${{\mathbf{X}}^{[i]}}(n)\in {{\mathbb{C}}^{{{A}_{i}}\times 1}}$ is the signal transmitted by transmitter $i$, ${{\mathbf{Y}}^{[j]}}(n)\in {{\mathbb{C}}^{{{B}_{j}}\times 1}}$  is the signal received by receiver $j$ and ${{\mathbf{Z}}^{[j]}}(n)\in {{\mathbb{C}}^{{{B}_{j}}\times 1}}$  denotes the additive Gaussian noise (AWGN) at receiver $j$. The average power at each transmitter is bounded by $\rho$ and the noise variance at all receivers is assumed to be equal to unity. ${{\mathbf{H}}^{[ji]}}(n)\in {{\mathbb{C}}^{{{B}_{j}}\times {{A}_{i}}}}$   represents the channel matrix from transmitter $i$ to receiver $j$ in time slot $n$. We assume that all channel coefficients values in different fading blocks are drawn from an i.i.d. continuous distribution and the absolute value of all the channel coefficients is bounded between a non-zero minimum value and a finite maximum value. Each receiver has a perfect estimate of its CSI, i.e., has perfect (global) CSIR. We ignore noise in this paper because, for linear beamforming schemes, noise does not affect sum-DoF.

Assuming that the error-free feedback links have  feedback delay of ${T}_{fb}$ time slots, transmitter $i\!\!\in\!\!\{1,2,...,M\}$ has access to local CSI $\mathbf{H}_{n-{{T}_{fb}}}^{[ji]}\!\!\!=\!\{{{\mathbf{H}}^{[ji]}}(1),{{\mathbf{H}}^{[ji]}}(2),...,{{\mathbf{H}}^{[ji]}}(n-{{T}_{fb}})\}$ up to time $n$ for receivers $j\!\!\in\!\!\{1,2,...,N\}$. We denote the local and delayed CSI matrix known to transmitter $i$ in time slot $n$ by $\mathbf{H}_{n-{{T}_{fb}}}^{i}\!\!\!\!=\!\!\{\mathbf{H}_{n-{{T}_{fb}}}^{[1i]},\mathbf{H}_{n-{{T}_{fb}}}^{[2i]},...,\mathbf{H}_{n-{{T}_{fb}}}^{[Ni]}\}$. Then, the signal transmitted by transmitter $i$ is generated as a function of the transmitted messages and the delayed and local CSIT, i.e., ${{\mathbf{X}}^{[i]}}(n)={{f}_{i}}({{W}^{[1i]}},{{W}^{[2i]}},...,{{W}^{[Ni]}},\mathbf{H}_{n-{{T}_{fb}}}^{i})$, where ${{f}_{i}}(*)$  represents the encoding function for transmitter $i$. 

\subsection{Block Fading and CSI Feedback Model}
Following the terminology of \cite{lee2014space}, we define an ideal block fading channels where the channel values remain invariant during the channel coherence time $T_c$  and change independently between blocks. Each transmitter is able to continuously track all variations in the channel changes since each receiver perfectly estimates CSI from different transmitters and sends it back to the corresponding transmitters every $T_c$  time slots periodically through error-free but delayed feedback links. 

We further assume that the feedback delay $T_{fb}$ is less than the channel coherence time, i.e., ${T_{fb}} < {T_c}$. An interesting fact about this CSI feedback model is that it allows transmitter $i$ to obtain the current CSI due to the channel invariance for every channel block. For example, as illustrated in Fig. 2, transmitter $i$ can access to the current CSI of the second channel block as well as the outdated CSI of the previous channel blocks in time slot 6. We provide a parameter namely the normalized CSI feedback delay to characterize CSI unit obsoleteness, i.e., $\lambda  = \frac{{{T_{fb}}}}{{{T_c}}}$. 
\begin{figure}[!t]
\centering
\includegraphics[width=3.5in]{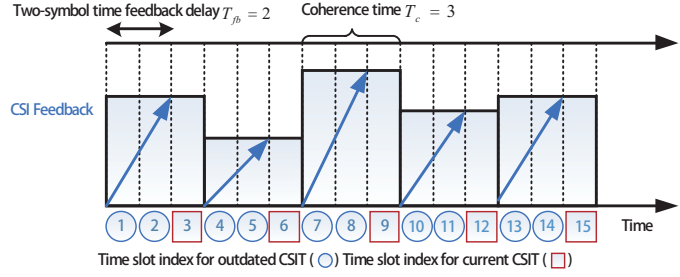}
\caption{Illustration of the ideal block fading channels.}
\label{fig.2}
\end{figure}

\subsection{Sum-DoF and CSI Feedback Delay Trade-Off}
Since the achievable data rate of the users depends on the normalized CSI feedback delay $\lambda$ and signal-to-noise ratio (SNR), we express it as a function of $\lambda$ and $SNR$ [25]. Specifically, for codewords spanning over $n$ channel uses, a rate of message $W^{[ji]}$, ${R_{ji}}(\lambda ,SNR) = \frac{{{{\log }_2}\left| {{W^{[ji]}}(\lambda ,SNR)} \right|}}{n}$, is achievable if the probability of error for $W^{[ji]}$ approaches zero as $n$ goes to infinity. The DoF of $W^{[ji]}$ is defined as
${d_{ji}}=\mathop {\lim }\nolimits_{SNR \to \infty } \frac{{{R_{ji}}(\lambda ,SNR)}}{{\log (SNR)}}$. Thus, the sum-DoF trade-off of the MIMO X network is given by $d_\Sigma ^X(M,N;\lambda ) = \sum\limits_{i,j} {{d_{ji}}}$.

\section{Main Results and Comparisons}
\subsection{Main Results}
The main results of this paper are presented in the following three theorems, and their proofs are provided in Sections IV, V and VI. We characterize three achievable sum-DoF regions each as a function of the normalized CSI feedback delay $\lambda$ for the two-user MIMO X channel, $K$-user SISO X network and ${M}\times{N}$ MISO user X network, respectively. 
\begin{theorem}
For the two-user MIMO X channel with $local$ CSIT, where each transmitter has $A$ antennas and each receiver has $B$ antennas, an achievable trade-off region between the sum-DoF and $\lambda$ is given as follows:
\begin{equation}
d_\Sigma ^{{X_L}}(2,2;\lambda ) = \left\{ {\begin{array}{*{20}{c}}
{{\textstyle{{4A} \over {{T_{AB}}}}},}\\
{a(A,B)\lambda  + b(A,B),}\\
{\min (2A,B),}
\end{array}} \right.\begin{array}{*{20}{c}}
{0 \le \lambda  \le {\textstyle{2 \over {{T_{AB}}}}},}\\
{{\textstyle{2 \over {{T_{AB}}}}} < \lambda  < 1,}\\
{\lambda  \ge 1.}
\end{array}
\end{equation}
where ${T_{AB}} \!\!= \!\!2 + \left\lceil {\frac{{2A - B}}{B}} \right\rceil$, $a(A,B) \!\!=\!\! {\textstyle{{4A - {T_{AB}}\min (2A,B)} \over {2 - {T_{AB}}}}}$ and $b(A,B) \!= {\textstyle{{2\min (2A,B) - 4A} \over {2 - {T_{AB}}}}}$.
\end{theorem}
\begin{theorem}
For the $K$-user SISO X network with $local$ CSIT, the achievable CSI feedback delay-DoF gain trade-off region is given by
\begin{equation}
d_\Sigma ^{{X_L}}(K,K;\lambda ) = \left\{ {\begin{array}{*{20}{c}}
{{\textstyle{{2(2K - 1)} \over {3K - 1}}},}\\
{ - {\textstyle{1 \over 3}}\lambda  + {\textstyle{4 \over 3}},}\\
{1,}
\end{array}} \right.\begin{array}{*{20}{c}}
{0 \le \lambda  \le {\textstyle{2 \over {3K - 1}}},}\\
{{\textstyle{2 \over {3K - 1}}} < \lambda  < 1,}\\
{\lambda  \ge 1.}
\end{array}
\end{equation}
\end{theorem}
\begin{theorem}
For the \textit{M}$\times$\textit{N} ($N\!\ge\!$ 3) user MISO X network with $local$ CSIT, where each transmitter has $A\!=\!N-1$  antennas and each receiver has a single antenna, an achievable trade-off region between the sum-DoF and $\lambda$  is given as follows:
\begin{equation}
d_\Sigma ^{{X_L}}(M,N;\lambda ) = \left\{ {\begin{array}{*{20}{c}}
{{\textstyle{{MN(N - 1)} \over {T_{MN}}}},}\\
{c(M,N)\lambda  + d(M,N),}\\
{1.}
\end{array}} \right.\begin{array}{*{20}{c}}
{0 \le \lambda  \le {\textstyle{2 \over {T_{MN}}}},}\\
{{\textstyle{2 \over {T_{MN}}}} < \lambda  < 1,}\\
{\lambda  \ge 1.}
\end{array}
\end{equation}
where $T_{MN}\!\!=\!\!M(N-1)+1$, $c(M,N)\! \!=\!\! {\textstyle{{1 - M{{(N - 1)}^2}} \over {M(N - 1) - 1}}}$ and $d(M,N) = {\textstyle{{MN(N - 1) - 2} \over {M(N - 1) - 1}}}$.
\end{theorem}

\textit{Remark 1 (Spatial Scale Invariance):} With the sum-DoF achieving transmission schemes discussed in the following content, the spatial scale invariance property proposed in \cite{Wang2011Subspace,Sun2012Degrees} is still valid with the temperately-delayed local CSIT assumption, i.e., if the number of antennas at each node is scaled by a common constant factor $q$, then the DoF of the network scale by the same factor. Therefore, using the scaled schemes proposed in Sections V and VI, $qd_\Sigma ^{{X_L}}(K,K;\frac{2}{{3K - 1}})$ and $qd_\Sigma ^{{X_L}}(M,N;\frac{2}{{M(N - 1) + 1}})$ are achievable in the $K$-user MIMO X network and \textit{M}$\times$\textit{N} user MIMO X network, respectively, where $d_\Sigma ^{{X_L}}(K,K;\frac{2}{{3K - 1}})$ and $d_\Sigma ^{{X_L}}(M,N;\frac{2}{{M(N - 1) + 1}})$ are given in Sections V and VI, respectively.
\begin{table}[]
\renewcommand\arraystretch{1}
\centering
\caption{Sum-DoFs of the two-user MIMO X channel under different CSIT assumptions.}
\begin{tabular*}{3.3in}{c|cccc}
\hline
Case No. &$d_{{\rm{STIA}}}^{{X_L}}$ &$d_{{\rm{GAK}}}^{{X_G}}$ &$d_{{\rm{IA}}}^{{X_F}}$&$d_{{\rm{VV}}}^{{X_N}}$\\
\hline
$2B \le A$&$\frac{{4A}}{{{T_{AB}}}}$&$\frac{4B}{3}$&$2B$&$B$\\
$B<A<2B$&$\frac{{4A}}{{{T_{AB}}}}$&$\frac{2B(A+2B)}{A+4B}$&$\min (2B,\frac{{4A}}{3})$&$B$\\
$\frac{3B}{4}<A\le B$&$\frac{{4A}}{{3}}$&$\frac{6B}{5}$&$\min (2A,\frac{{4B}}{3})$&$B$\\
$\frac{B}{2}<A \le \frac{3B}{4}$&$\frac{{4A}}{{3}}$&$\frac{4AB}{2A+B}$&$\min (2A,\frac{{4B}}{3})$&$B$\\
$A\le \frac{B}{2}$&$2A$&$2A$&$2A$&$2A$\\
\hline
\end{tabular*}
\end{table}

\subsection{Comparisons of Achievable Trade-offs}
To reveal the impact of the distributed CSIT on the sum-DoF of the two-user MIMO X channel, we first compare the achievable sum-DoF under local and temperately-delayed CSIT with the achievable sum-DoF  using GAK scheme in \cite{Ghasemi2012On}, where global and completely-delayed CSIT is considered. The comparison results are summarized in Table I along with other achievable regions with full CSIT or without CSIT. For simplicity, we denote $d_{{\rm{STIA}}}^{{X_L}}$, $d_{{\rm{GAK}}}^{{X_G}}$, $d_{{\rm{IA}}}^{{X_F}}$ and $d_{{\rm{VV}}}^{{X_N}}$ as the sum-DoFs achievable by our proposed STIA scheme, the GAK scheme in [13], the interference alignment (IA) scheme in [2], and the VV scheme in [19], respectively. From Table I, we have the following observations: 
\begin{itemize}
\item[$\bullet$] For ${2B}\le{A}$, local CSIT contributes to attain better sum-DoFs than those obtained under global and no CSIT assumptions. For example, when $A=5$ and $B=2$, the proposed method achieves $\frac{10}{3}$  sum-DoF that significantly exceeds the $\frac{8}{3}$  sum-DoF under the global and completely-delayed CSIT case and 2 sum-DoF under the no CSIT case, where the sum-DoF with full CSIT is 4.
\item[$\bullet$] For $\frac{{3B}}{4}\!\!<\!\!A\!\!<\!\!{2B}$, local CSIT improves the sum-DoF compared to the no CSIT case. Another interesting finding is that the achievable sum-DoF with local CSIT may be higher than the global CSIT case on certain configurations. For instance, when $A=5$ and $B=3$, our achievable sum-DoF is 4 which is strictly better than the $\frac{66}{17}$ sum-DoF with global CSIT and 3 sum-DoF with no CSIT, respectively, while the sum-DoF under the full CSIT assumption is 6. It is also remarkable that the sum-DoF of 4 is greater than the sum-DoF of $\frac{90}{23}$ achieved by a linear coding strategy in [18]. Another similar case can be $A=10$ and $B=11$, where achievable sum-DoF for local CSIT is $\frac{40}{3}$  while $\frac{66}{5}$  and 10 sum-DoFs are achievable for the global CSIT case and the no CSIT case, respectively.
\item[$\bullet$] For $\frac{{3B}}{4}\!\!<\!\!A$, the achievable result under the local CSIT assumption lies strictly between the regions with full and no CSIT.
\end{itemize}

We note that, comparing with the delayed CSIT and no CSIT cases, distributed CSIT still contributes to increase the DoF performance. To shed further light on how CSI feedback delay affects the sum-DoF, we establish another trade-off region for the two-user MIMO X channel with global and delayed CSIT as follows:

\begin{corollary}
For the  two-user MIMO X channel  with $global$ CSIT, where each transmitter has $A$ antennas and each receiver has $B$ antennas (${2B}\le{A}$), an achievable trade-off region between the sum-DoF and $\lambda$ is given by
\begin{equation}
d_\Sigma ^{{X_G}}(2,2;\lambda ) = \left\{ {\begin{array}{*{20}{c}}
{{\textstyle{{4A} \over {{T_{AB}}}}},}\\
{e(A,B)\lambda  + f(A,B),}\\
{\frac{{4B}}{3},}
\end{array}} \right.\begin{array}{*{20}{c}}
{0 \le \lambda  \le {\textstyle{2 \over {{T_{AB}}}}},}\\
{{\textstyle{2 \over {{T_{AB}}}}} < \lambda  < 1,}\\
{\lambda  \ge 1.}
\end{array}
\end{equation}
where ${T_{AB}} = 2 + \left\lceil {\frac{{2A - B}}{B}} \right\rceil $, $e(A,B) = {\textstyle{{4B{T_{AB}} - 12A} \over {3({T_{AB}} - 2)}}}$ and $f(A,B) = {\textstyle{{12A - 8B} \over {3({T_{AB}} - 2)}}}$.
\end{corollary}

\textit{Proof:} That $\frac{4B}{3}$ is the achievable sum-DoF found for this channel follows from the corresponding GAK scheme for the ${2B}\le{A}$ antenna configuration in \cite{Ghasemi2012On}. Note that an achievable result for the ${2B}\le{A}$ antenna configuration with global CSIT is also an achievable result for our setting because global CSIT becomes available for the completely delayed regime,  $\lambda \ge 1$. Thus, with the achievable sum-DoF of $d_\Sigma ^{{X_G}}(2,2;\lambda ) = \frac{{4A}}{{{T_{AB}}}}$ derived from Theorem 1 for $0\le\lambda\le\frac{2}{T_{AB}}$, the achievability of the new trade-off region between the sum-DoF and the CSI feedback delay $\lambda$ can be spread over global CSIT setting, where a time-sharing technique between these two schemes is used to achieve any points in the line connecting two points between $d_\Sigma ^{{X_G}}(2,2;\frac{2}{{{T_{AB}}}})$ and $d_\Sigma ^{{X_G}}(2,2;1)$. It is remarkable that similar results can be obtained for the other antenna configurations. $\quad \quad \quad \quad \quad \quad\quad \quad\quad\quad\quad\quad\quad\quad\quad\quad\quad \quad\quad \quad\quad\quad\quad\quad\quad\quad\quad\quad\quad\quad\quad\quad\quad\Box$

We next compare it with other regions achieved by different methods when $A=5$, and $B=2$. As illustrated in Fig. 3, the IA-TDMA region of $d_{\rm{IA - TDMA}}^{{X_G}}(2,2;\lambda ) =  - 2\lambda  + 4$ can be achievable with global CSIT for $0\le\lambda\le1$ by using a time sharing technique between IA and TDMA scheme. A same argument is applied for the other regions. Note that the STIA-TDMA region coincides with the IA-TDMA region for $\frac{2}{6} \le \lambda  \le 1$, which implies that a time sharing technique between STIA and TDMA can provide a tight bound for this antenna configuration. It is also notable that global CSIT allows to attain a higher trade-off region for the STIA-GAK scheme for $\frac{2}{6} \le \lambda  \le 1$, compared to the STIA-TDMA region where local CSIT is applied. Whereas, only local CSIT is enough for the case of $\lambda  \le \frac{2}{6}$ because global CSIT does not improve the sum-DoF here. 
\begin{figure}[]
\centering
\includegraphics[width=2.4in, angle=270]{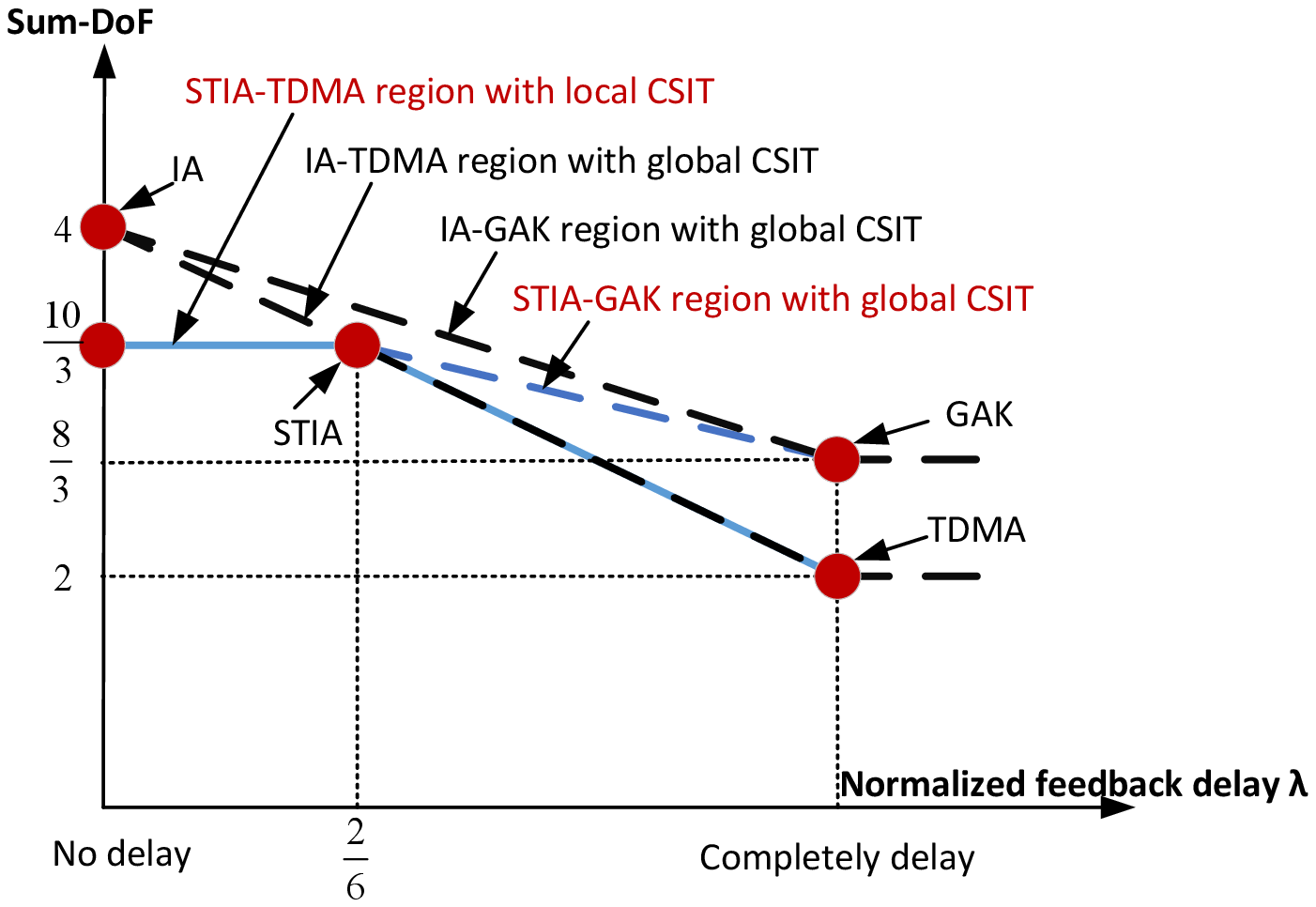}
\caption{Illustration of trade-offs for the two-user MIMO X channel when $A=5$ and $B=2$.}
\label{fig.3}
\end{figure}
\begin{figure}[]
\centering
\includegraphics[width=2.5in, angle=270]{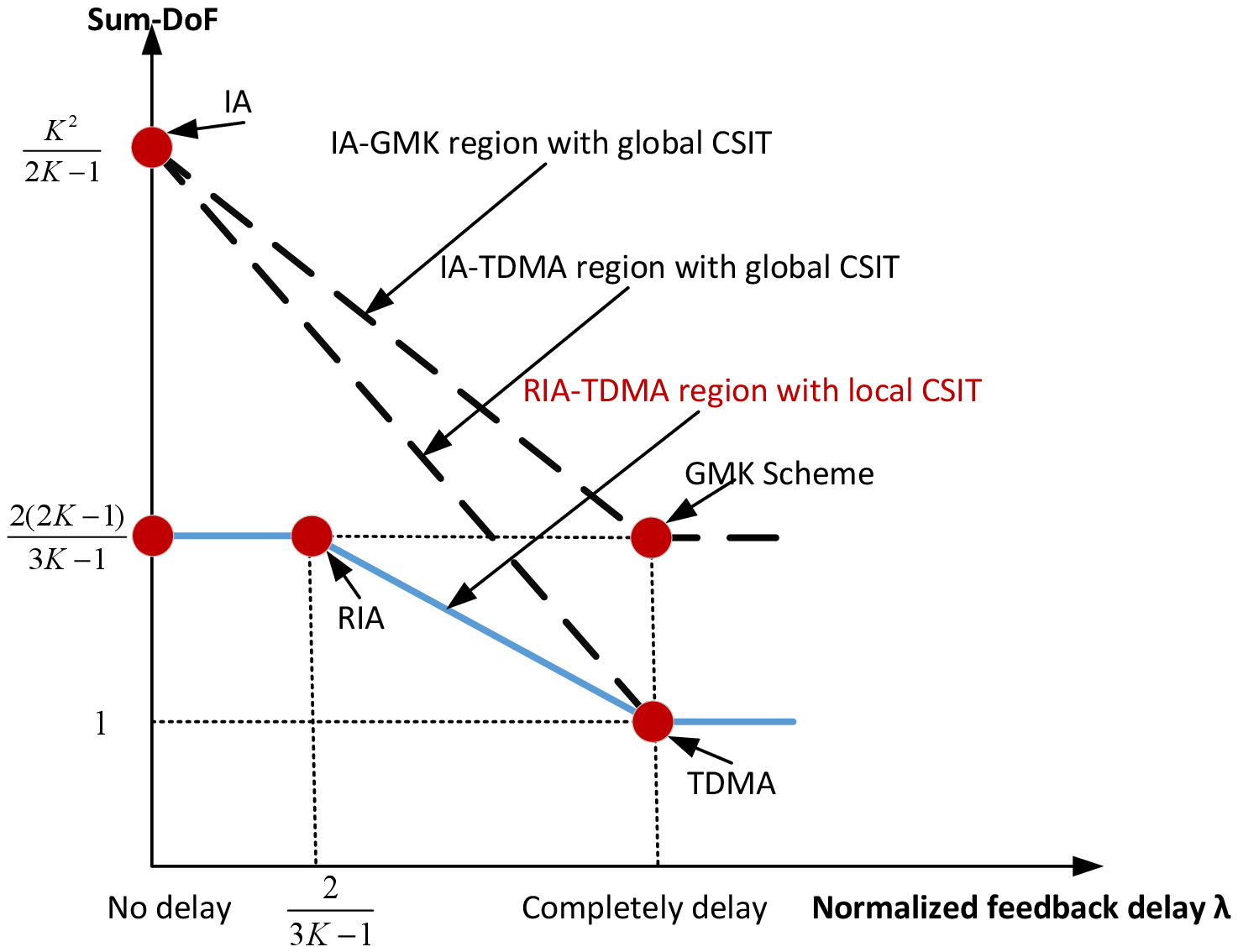}
\caption{Illustration of trade-offs for the $K$-user SISO X network.}
\label{fig.4}
\end{figure}

Similarly, as illustrated in Fig. 4, a comparison between the achievable trade-off region in Theorem 2 and other regions achievable with global CSIT (GMK scheme in [11]) is given. We show that a $K$-user SISO X network only with local CSIT can achieve more than 1 sum-DoF. We further compare the achievable trade-off region in Theorem 3 with the region under the full CSIT assumption. As shown in Fig. 5, in the context of \textit{M}$\times$\textit{N} user MISO X network, the proposed method with local CSIT allows to attain a higher trade-off region between the sum-DoF and CSI feedback delay than the IA scheme does when the delay of the CSI feedback is not severe.

\begin{figure}[!t]
\centering
\includegraphics[width=2.5in, angle=270]{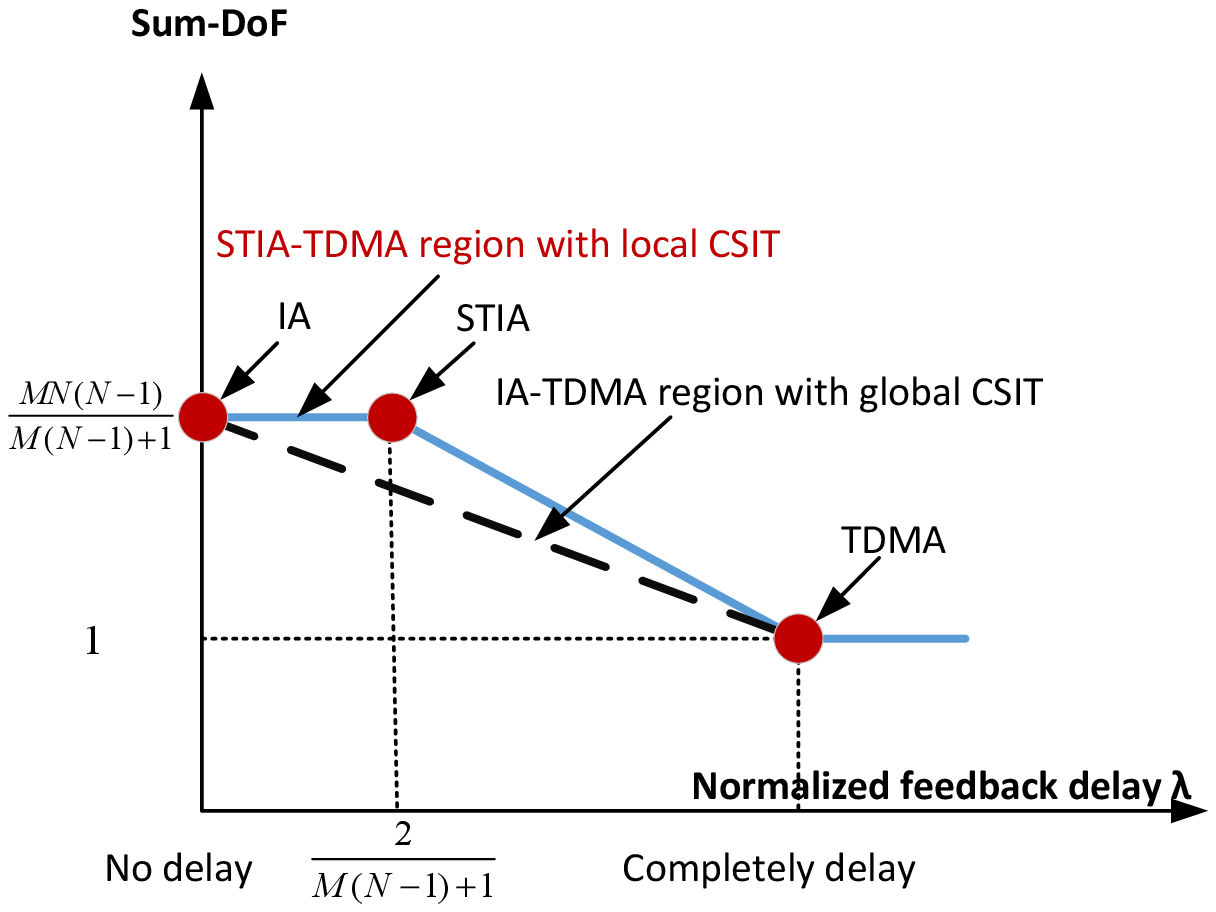}
\caption{Illustration of trade-offs for \textit{M}$\times$\textit{N} user MISO X network.}
\label{fig.5}
\end{figure}
\section{Transmission Scheme Achieving the Sum-DoF in Theorem 1}
In this section, we specify the transmission scheme that achieves the sum-DoF in Theorem 1. To better explain our idea, we start with the two-user MISO X channel under the local and temperately-delayed CSIT assumption and then give a general proof to the theorem. 

\subsection{Two-User MISO X Channel}
Consider the two-user MISO X channel where each transmitter $i \in \{1,2\}$ has two antennas and each receiver $j \in \{1,2\}$  has a single antenna. We focus on the special case of $\lambda=\frac{2}{5}$, i.e., each transmitter has access to current CSIT over three-fifths of the channel coherence time. We show that $\frac{8}{5}$  sum-DoF is achievable, i.e., 8 independent information symbols will be transmitted over 5 channel uses. In particular, we select $n\in \{1,6,13,18,23\}$  five time slots belonging to different channel coherence blocks. Note that all channel coefficients values are drawn from an i.i.d. continuous distribution. We refer to $\bf{u,v}$ as symbol vectors intended for receiver 1 and 2, respectively. The proposed transmission scheme involves two phases.

\textit{Phase one:} This phase takes two time slots, i.e.,  $n\in \{1,6\}$. In time slot 1, each transmitter sends a two-symbol vector intended for receiver 1, i.e., 
\begin{equation}
{{\bf{X}}^{[1]}}(1) = {{\bf{u}}^{[1]}},
{{\bf{X}}^{[2]}}(1) = {{\bf{u}}^{[2]}}, 
\end{equation}
where ${{\bf{u}}^{[1]}} \!\!=\! {[u_1^{[1]},u_2^{[1]}]^T}$  and ${{\bf{u}}^{[2]}} \!\!=\! {[u_1^{[2]},u_2^{[2]}]^T}$. Then, at receiver $j$, for $j \in \{1,2\}$, we have
\begin{equation}
{y^{[j]}}(1) = {{\bf{h}}^{[j1]}}(1){{\bf{u}}^{[1]}} + {{\bf{h}}^{[j2]}}(1){{\bf{u}}^{[2]}},
\end{equation}
where ${{\bf{h}}^{[ji]}}(1) \!\in\! {\mathbb{C}^{1 \times 2}}$ denotes the channel vector from transmitter $i$ to receiver $j$, for $i,j \in \{ 1,2\}$.

In time slot 6, each transmitter sends the two-symbol vector intended for receiver 2, i.e., 
\begin{equation}
{{\bf{X}}^{[1]}}(6) = {{\bf{v}}^{[1]}},
{{\bf{X}}^{[2]}}(6) = {{\bf{v}}^{[2]}},
\end{equation}
where ${{\bf{v}}^{[1]}} = {[v_1^{[1]},v_2^{[1]}]^T}$ and ${{\bf{v}}^{[2]}} = {[v_1^{[2]},v_2^{[2]}]^T}$. Therefore, at receiver $j$, for $j \in \{1,2\}$, we have
\begin{equation}
{y^{[j]}}(6) = {{\bf{h}}^{[j1]}}(6){{\bf{v}}^{[1]}} + {{\bf{h}}^{[j2]}}(6){{\bf{v}}^{[2]}}.
\end{equation}

\textit{Phase two:} This phase takes three time slots, i.e.,  $n\in\{13,18,23\}$. In each time slot of this phase, each transmitter sends a superposition of two-symbol vectors they ever sent after precoding, i.e., 
\begin{equation}
{{\bf{X}}^{[1]}}(n) = {\bf{V}}_1^{[1]}(n){{\bf{u}}^{[1]}} + {\bf{V}}_2^{[1]}(n){{\bf{v}}^{[1]}},
{{\bf{X}}^{[2]}}(n) = {\bf{V}}_1^{[2]}(n){{\bf{u}}^{[2]}} + {\bf{V}}_2^{[2]}(n){{\bf{v}}^{[2]}},
\end{equation}
where ${\bf{V}}_j^{[i]}(n) \in {\mathbb{C}^{2 \times 2}}$ denotes the precoding matrix used for carrying the same symbol vectors ${{\bf{u}}^{[i]}}$  and ${{\bf{v}}^{[i]}}$ in time slot $n$, where  $i,j\in\{1,2\}$ and $n\in\{13,18,23\}$. The main idea for designing the precoding matrix is to ensure that each receiver exactly sees the aligned interference shape that it previously obtained by exploiting both current and outdated CSI. Recall that each receiver has obtained a linear combination of desired symbols as well as a linear combination of undesired symbols by the end of phase one. Therefore, transmitter 1 constructs the precoding matrices ${\bf{V}}_1^{[1]}(n)$  and ${\bf{V}}_2^{[1]}(n)$  to satisfy
\begin{equation}
{{\bf{h}}^{[21]}}(n){\bf{V}}_1^{[1]}(n) = {{\bf{h}}^{[21]}}(1),
{{\bf{h}}^{[11]}}(n){\bf{V}}_2^{[1]}(n) = {{\bf{h}}^{[11]}}(6).
\end{equation}

Since the channel matrix is a vector, matrix inversion here is unavailable. Recall that the channel values do not change over the same channel block, which implies that, in time slot $n$ ($n\in\{13,18,23\}$), each transmitter is able to access current CSI, i.e.,  ${{\bf{h}}^{[ji]}}(n) = {{\bf{h}}^{[ji]}}(n - 2)$. With the help of current and delayed CSI, however, we can find the special precoding matrices, of which the back-diagonal elements are zeroes, to satisfy the equations, i.e., 
\begin{equation}
{\bf{V}}_1^{[1]}(n) = \left[ {\begin{array}{*{20}{c}}
{\frac{{h_1^{[21]}(1)}}{{h_1^{[21]}(n - 2)}}}&0\\
0&{\frac{{h_2^{[21]}(1)}}{{h_2^{[21]}(n - 2)}}}
\end{array}} \right],
{\bf{V}}_2^{[1]}(n) = \left[ {\begin{array}{*{20}{c}}
{\frac{{h_1^{[11]}(6)}}{{h_1^{[11]}(n - 2)}}}&0\\
0&{\frac{{h_2^{[11]}(6)}}{{h_2^{[11]}(n - 2)}}}
\end{array}} \right].
\end{equation}

Similarly, transmitter 2 constructs the precoding matrices ${\bf{V}}_1^{[2]}(n)$  and ${\bf{V}}_2^{[2]}(n)$  carrying the two-symbol vectors, ${\bf{u}}^{[2]}$ and  ${\bf{v}}^{[2]}$, to satisfy
\begin{equation}
{{\bf{h}}^{[22]}}(n){\bf{V}}_1^{[2]}(n) = {{\bf{h}}^{[22]}}(1),
{{\bf{h}}^{[12]}}(n){\bf{V}}_2^{[2]}(n) = {{\bf{h}}^{[12]}}(6).
\end{equation}
where the precoding matrices can be written as
\begin{equation}
{\bf{V}}_1^{[2]}(n) = \left[ {\begin{array}{*{20}{c}}
{\frac{{h_1^{[22]}(1)}}{{h_1^{[22]}(n - 2)}}}&0\\
0&{\frac{{h_2^{[22]}(1)}}{{h_2^{[22]}(n - 2)}}}
\end{array}} \right],
{\bf{V}}_2^{[2]}(n) = \left[ {\begin{array}{*{20}{c}}
{\frac{{h_1^{[12]}(6)}}{{h_1^{[12]}(n - 2)}}}&0\\
0&{\frac{{h_2^{[12]}(6)}}{{h_2^{[12]}(n - 2)}}}
\end{array}} \right].
\end{equation}

Thus, the received signals at receiver 1 and 2 in time slot $n$ are given by
\begin{eqnarray}
{y^{[1]}}(n) &=& {{\bf{h}}^{[11]}}(n){{\bf{X}}^{[1]}}(n) + {{\bf{h}}^{[12]}}(n){{\bf{X}}^{[2]}}(n),\nonumber \\
             & = & {{\bf{h}}^{[11]}}(n){\bf{V}}_1^{[1]}(n){{\bf{u}}^{[1]}} + {{\bf{h}}^{[12]}}(n){\bf{V}}_1^{[2]}(n){{\bf{u}}^{[2]}}+\underbrace { {{\bf{h}}^{[11]}}(6){{\bf{v}}^{[1]}} + {{\bf{h}}^{[12]}}(6){{\bf{v}}^{[2]}}}_{{y^{[1]}}(6)}.\\
{y^{[2]}}(n) &=& {{\bf{h}}^{[21]}}(n){{\bf{X}}^{[1]}}(n) + {{\bf{h}}^{[22]}}(n){{\bf{X}}^{[2]}}(n),\nonumber \\
             &=& {{\bf{h}}^{[21]}}(n){\bf{V}}_2^{[1]}(n){{\bf{v}}^{[1]}} + {{\bf{h}}^{[22]}}(n){\bf{V}}_2^{[2]}(n){{\bf{v}}^{[2]}}+\underbrace { {{\bf{h}}^{[21]}}(1){{\bf{u}}^{[1]}} + {{\bf{h}}^{[22]}}(1){{\bf{u}}^{[2]}}}_{{y^{[2]}}(1)}.
\end{eqnarray}

Next we explain how every receiver has enough information to recover its desired symbols. Consider receiver 1, it obtains three fresh linear combinations containing of four desired symbols, $\{ u_1^{[1]},u_2^{[1]},u_1^{[2]},u_2^{[2]}\}$, at the end of phase two, by performing the interference cancellation, i.e.,  ${y^{[1]}}(n) - {y^{[1]}}(6)$. Therefore, there are four different equations in total and the concatenated input-output relationship is given as
\begin{equation}
\left[ {\begin{array}{*{20}{c}}
{{y^{[1]}}(1)}\\
{{y^{[1]}}(13)\! -\! {y^{[1]}}(6)}\\
{{y^{[1]}}(18) \!- \!{y^{[1]}}(6)}\\
{{y^{[1]}}(23)\! - \!{y^{[1]}}(6)}
\end{array}} \right] \!\!=\!\! \underbrace {\left[ {\begin{array}{*{20}{c}}
{h_1^{[11]}(1)}&{h_2^{[11]}(1)}&{h_1^{[12]}(1)}&{h_2^{[12]}(1)}\\
{h_1^{[11]}(13)\frac{{h_1^{[21]}(1)}}{{h_1^{[21]}(13)}}}&{h_2^{[11]}(13)\frac{{h_2^{[21]}(1)}}{{h_2^{[21]}(13)}}}&{h_1^{[12]}(13)\frac{{h_1^{[22]}(1)}}{{h_1^{[22]}(13)}}}&{h_2^{[12]}(13)\frac{{h_2^{[22]}(1)}}{{h_2^{[22]}(13)}}}\\
{h_1^{[11]}(18)\frac{{h_1^{[21]}(1)}}{{h_1^{[21]}(18)}}}&{h_2^{[11]}(18)\frac{{h_2^{[21]}(1)}}{{h_2^{[21]}(18)}}}&{h_1^{[12]}(18)\frac{{h_1^{[22]}(1)}}{{h_1^{[22]}(18)}}}&{h_2^{[12]}(18)\frac{{h_2^{[22]}(1)}}{{h_2^{[22]}(18)}}}\\
{h_1^{[11]}(23)\frac{{h_1^{[21]}(1)}}{{h_1^{[21]}(23)}}}&{h_2^{[11]}(23)\frac{{h_2^{[21]}(1)}}{{h_2^{[21]}(23)}}}&{h_1^{[12]}(23)\frac{{h_1^{[22]}(1)}}{{h_1^{[22]}(23)}}}&{h_2^{[12]}(23)\frac{{h_2^{[22]}(1)}}{{h_2^{[22]}(23)}}}
\end{array}} \right]}_{{{{\bf{\hat H}}}_1}}\left[ {\begin{array}{*{20}{c}}
{u_1^{[1]}}\\
{u_2^{[1]}}\\
{u_1^{[2]}}\\
{u_2^{[2]}}
\end{array}} \right].
\end{equation}

Recall that all precoding matrices ${\bf{V}}_j^{[i]}(n)$, $i,j \in \{ 1,2\}$, are independently generated regardless of the direct channel ${{\bf{h}}^{[ji]}}(n)$  for $n \in \{ 1,2,...,n - 2\}$. Further, it is assumed that all channel values are drawn from an i.i.d. continuous distribution across time and space. Therefore, the effective channel ${{\bf{\hat H}}_1}$ for receiver 1 has a full rank almost surely, i.e., rank(${{\bf{\hat H}}_1}$)=4. Consequently, we see receiver 1 has obtained four linear independent combinations for four desired symbols. By symmetry, receiver 2 operates in a similar fashion, which implies that a total DoF of $\frac{8}{5}$  is achievable.

\textit{Remark 2 (Decomposability of the Two-User MISO X Channel):} We refer to decomposability as an independent processing at each antenna, essentially splitting a multiple antenna node into multiple independent single antenna nodes. Recall that the optimal sum-DoF of $\frac{{2K}}{{K + 1}}$  is achievable almost surely for the $K\times2$  SISO X channel with local CSIT \cite{Lee2014Distributed}. We argue that such a two-user $2\times1$ vector MISO X channel can be broken up into a $4\times2$  SISO X channel where the same sum-DoF of $\frac{8}{5}$  is achievable, without joint processing among collocated antennas at any node. The new result stated above is strictly better than the best known lower bound under feedback and delayed CSI assumption in \cite{Tandon2012On}. More generally, if we select and integrate an arbitrary number of nodes among $K$ in \cite{Lee2014Distributed} to be the transmitter 1 with $A_1$  antennas and the rest part to be the transmitter 2 with $A_2$  antennas where $A_1+A_2=K$, the two-user MISO X channel with such an asymmetric antenna configuration can achieve $\frac{{2({A_1} + {A_2})}}{{{A_1} + {A_2} + 1}}$ sum-DoF, almost surely. And one can easily prove that the precoding matrices  ${\bf{V}}_j^{[i]} \in {\mathbb{C}^{{A_i} \times {A_i}}}$ for $i,j\in\{1,2\}$  are diagonal. Without loss of generality, another extension can be inferred subsequently where the $M\times2$ $(M\ge 3)$  user MISO X channel with temperately-delayed local CSI can achieve  $\frac{{2({A_1} + {A_2} + \cdots + {A_M})}}{{{A_1} + {A_2} + \cdots + {A_M} + 1}}$ sum-DoF, almost surely. 

\subsection{Two-User MIMO X Channel: Proof of Theorem 1}
We refer to the case where $\lambda \ge 1$ as the completely delayed local CSIT point. By a TDMA transmission method for this case, one can easily infer that $d_\Sigma ^{{X_L}}(2,2;1) = \min (2A,B)$  is achievable [19]. Hence, by time sharing between the proposed STIA scheme and a TDMA method, we can obtain the points connecting two points $d_\Sigma ^{{X_L}}(2,2;\frac{2}{{{T_{AB}}}}) = \frac{{4A}}{{{T_{AB}}}}$  and  $d_\Sigma ^{{X_L}}(2,2;1) = \min (2A,B)$, where we take the sum-DoF as a linear equation of the CSI feedback delay  $\lambda$. Thus, we concentrate on the proof of the point $d_\Sigma ^{{X_L}}(2,2;\frac{2}{{{T_{AB}}}}) = \frac{{4A}}{{{T_{AB}}}}$  by considering each of the three cases separately as follows:

a)$B\le A$. 

In this case, we interpret the transmission method selecting $T_{AB}$  channel uses while the normalized CSI feedback delay is $\lambda = \frac{2}{T_{AB}}$, i.e., $4A$ information symbols are delivered over $T_{AB}$  channel uses. Consider  $n+T_{AB}-1$ channel blocks comprising of a total of ${T_{AB}}({n+T_{AB}-1})$  time slots so that each block has $T_{AB}$  time slots, i.e., $T_{c}$=$T_{AB}$. We define ${S_t} = \{ 1,2,...,{T_{AB}}(n + {T_{AB}} - 1)\}$  as a set of time slots for transmission. Since we assume that the normalized CSI feedback delay is  $\lambda = \frac{2}{T_{AB}}$, the total time slots set can be divided into two subsets, $S_c$ with $\left| {{S_c}} \right| = ({T_{AB}} - 2)(n + {T_{AB}} - 1)$  and $S_d$  with $\left| {{S_d}} \right| = 2(n + {T_{AB}} - 1)$. Here,  $S_c$ denotes the set of time slots when the transmitter is able to access both current and delayed CSI, and  $S_d$ represents time slot set corresponding to the case where the transmitter has delayed CSI only. Further, we define $n$ time slot sets,  $\{ {I_1},{I_2},...,{I_n}\}$, each of which has $T_{AB}$  elements for applying the STIA scheme, i.e., ${I_l} = \{ {t_{l,1}},{t_{l,2}},...,{t_{l,{T_{AB}}}}\}$, where  $l\in\{1,2,...,n\}$, $\{ {t_{l,1}},{t_{l,2}}\}  \in {S_d}$, and ${t_{l,k}} \in {S_c}$ for  $k \in \{ 3,4,...,{T_{AB}}\}$. Note that any two time slots of  $I_l$ belong to different channel blocks. For example, when $T_{AB}=5$  and $n=3$, a total of 7 channel blocks comprising of 35 time resources can definitely provide three index sets for the proposed transmission method, i.e., ${I_1} = \{ 1,6,13,18,23\}$, ${I_2} = \{ 2,7,14,19,24\}$, and  ${I_3} = \{ 3,8,15,20,25\}$. Next we prove the achievability of sum-DoF for each time slot set  $I_l$ and we omit the index $l$  for simplicity, i.e., ${I_l} = \{ {t_1},{t_2},...,{t_{{T_{AB}}}}\}$. The proposed transmission scheme involves two phases.

\textit{Phase one:} It consists of two time slots belonging to  $\{t_1,t_2\}$. In time slot  $t_1$, each transmitter sends a $A$-symbol vector intended for receiver 1, i.e., ${{\bf{X}}^{[1]}}({t_1}) = {{\bf{u}}^{[1]}}$,  ${{\bf{X}}^{[2]}}({t_1}) = {{\bf{u}}^{[2]}}$, where ${{\bf{u}}^{[1]}} = {\left[ {u_1^{[1]},...,u_A^{[1]}} \right]^T}$  and ${{\bf{u}}^{[2]}} = {\left[ {u_1^{[2]},...,u_A^{[2]}} \right]^T}$  are the $A$-symbol vectors from transmitter 1 and 2, respectively. In time slot  $t_2$, each transmitter sends the $A$-symbol vector intended for receiver 2, i.e., ${{\bf{X}}^{[1]}}({t_2}) = {{\bf{v}}^{[1]}}$,  ${{\bf{X}}^{[2]}}({t_2}) = {{\bf{v}}^{[2]}}$, where ${{\bf{v}}^{[1]}} = {\left[{v_1^{[1]},...,v_A^{[1]}} \right]^T}$ and ${{\bf{v}}^{[2]}} = {\left[{v_1^{[2]},...,v_A^{[2]}} \right]^T}$. Note that each transmitter sends symbol vectors without a precoding technique in this phase for lack of channel knowledge. As a result, each receiver obtains $B$ linear independent combinations of $2A$ desired symbols, while overhearing $B$ linear independent combinations of $2A$ undesired symbols as follows
\begin{equation}
{{\bf{Y}}^{[j]}}({t_1}) = {{\bf{H}}^{[j1]}}({t_1}){{\bf{u}}^{[1]}} + {{\bf{H}}^{[j2]}}({t_1}){{\bf{u}}^{[2]}},
{{\bf{Y}}^{[j]}}({t_2}) = {{\bf{H}}^{[j1]}}({t_2}){{\bf{v}}^{[1]}} + {{\bf{H}}^{[j2]}}({t_2}){{\bf{v}}^{[2]}},
\end{equation}
where ${{\bf{H}}^{[ji]}}({t_1}) \in {\mathbb{C}^{B \times A}}$ denotes the channel matrix from transmitter $i$ to receiver $j$, $i,j\in\{1,2\}$.

\textit{Phase two:} Phase two consists of the rest time slots of $I_l$, i.e., $\{ {t_3},{t_4},...,{t_{{T_{AB}}}}\}$. Recall that transmitter $i$ with the set of delayed CSIT ${\bf{H}}_{n - 2}^i = \{ {\bf{H}}_{n - 2}^{[1i]},{\bf{H}}_{n - 2}^{[2i]}\}$  is able to access current CSI in phase two. We seek the possibility of aligning interference so that each receiver can obtain $B$ more linear independent combinations of desired symbols per time slot during phase two, by performing interference cancellation. Since  $B\le{A}$, each receiver needs $2A-B$ additional linear independent combinations of desired symbols. In that way, a number of $\left\lceil {\frac{{2A - B}}{B}} \right\rceil$ time slots are required for phase two, i.e., ${T_{AB}} - 2 = \left\lceil {\frac{{2A - B}}{B}} \right\rceil$. Thus, in each time slot of $n \in \{ {t_3},{t_4},...,{t_{{T_{AB}}}}\}$, two transmitters repeatedly multicast a superposition of the $A$-symbol vectors they ever sent after precoding in a distributed manner such that receiver 1 and receiver 2 observe the same interference symbols, respectively. Therefore, we construct the transmit vectors in time slot $n$ as 
\begin{equation}
{{\bf{X}}^{[i]}}(n) = {\bf{V}}_1^{[i]}(n){{\bf{u}}^{[i]}} + {\bf{V}}_2^{[i]}(n){{\bf{v}}^{[i]}},
\end{equation}
where $i \in \{ 1,2\}$, $n \in \{ {t_3},{t_4},...,{t_{{T_{AB}}}}\}$.  ${\bf{V}}_1^{[i]}(n),{\bf{V}}_2^{[i]}(n) \in {\mathbb{C}^{A \times A}}$ represent the precoding matrices generated at transmitter  $i$. As a result, receiver  $j$, for  $j\in\{1,2\}$, obtains
\begin{eqnarray}
\!\!\!\!{{\bf{Y}}^{[j]}}(n) \!\! &&=\! {{\bf{H}}^{[j1]}}(n){{\bf{X}}^{[1]}}(n) + {{\bf{H}}^{[j2]}}(n){{\bf{X}}^{[2]}}(n),\nonumber \\
 \!\! &&=\! {{\bf{H}}^{[j1]}}(n){\bf{V}}_1^{[1]}(n){{\bf{u}}^{[1]}} \!\!+\! {{\bf{H}}^{[j2]}}(n){\bf{V}}_1^{[2]}(n){{\bf{u}}^{[2]}}\!\!+ \!{{\bf{H}}^{[j1]}}(n){\bf{V}}_2^{[1]}(n){{\bf{v}}^{[1]}} \!\!+\! {{\bf{H}}^{[j2]}}(n){\bf{V}}_2^{[2]}(n){{\bf{v}}^{[2]}}. 
\end{eqnarray}
Note that the inverse of the channel matrix is nonexistent for  $A \ne B$. We now proceed to characterize the precoding matrices via an elegant way called \textit{Cyclic Zero-padding} and describe how every receiver performs interference cancellation. The precise construction of precoding matrices can be found in Appendix I.

Consider receiver 1, to ensure that receiver 1 can obtain $B$ more linear independent combinations of desired symbols per time slot during phase two, we construct ${\bf{V}}_j^{[i]}(n)$, $i\in\{1,2\}$, $j\in\{2\}$  to satisfy
\begin{eqnarray}
{{\bf{H}}^{[11]}}(n){\bf{V}}_2^{[1]}(n) = {{\bf{H}}^{[11]}}({t_2}),
{{\bf{H}}^{[12]}}(n){\bf{V}}_2^{[2]}(n) = {{\bf{H}}^{[12]}}({t_2}).
\end{eqnarray}
As shown in the Appendix I, we can obtain ${\bf{V}}_2^{[1]}(n)$ and  ${\bf{V}}_2^{[2]}(n)$ by \textit{Cyclic Zero-padding} and these two precoding matrices are full rank, almost surely. Thus, let ${{\bf{Y}}^{[1]}}(n)$ subtract ${{\bf{Y}}^{[1]}}({t_2})$, we have
\begin{equation}
{{\bf{Y}}^{[1]}}(n) - {{\bf{Y}}^{[1]}}({t_2}) = {{\bf{H}}^{[11]}}(n){\bf{V}}_1^{[1]}(n){{\bf{u}}^{[1]}} + {{\bf{H}}^{[12]}}(n){\bf{V}}_1^{[2]}(n){{\bf{u}}^{[2]}}.
\end{equation}
Likewise, for receiver 2, we construct  ${\bf{V}}_j^{[i]}(n)$, $i \in \{ 1,2\} ,j \in \{ 1\}$, to satisfy
\begin{eqnarray}
{{\bf{H}}^{[21]}}(n){\bf{V}}_1^{[1]}(n) = {{\bf{H}}^{[21]}}({t_1}),
{{\bf{H}}^{[22]}}(n){\bf{V}}_1^{[2]}(n) = {{\bf{H}}^{[22]}}({t_1}).
\end{eqnarray}
Let  ${{\bf{Y}}^{[2]}}(n)$ subtract ${{\bf{Y}}^{[2]}}({t_1})$, we have
\begin{equation}
{{\bf{Y}}^{[2]}}(n) - {{\bf{Y}}^{[2]}}({t_1}) = {{\bf{H}}^{[21]}}(n){\bf{V}}_2^{[1]}(n){{\bf{v}}^{[1]}} + {{\bf{H}}^{[22]}}(n){\bf{V}}_2^{[2]}(n){{\bf{v}}^{[2]}}.
\end{equation}
At the end of this phase, receiver 1 obtains a system of linear equations as
\begin{equation}
\left[ {\begin{array}{*{20}{c}}
{{{\bf{Y}}^{[1]}}({t_1})}\\
{{{\bf{Y}}^{[1]}}({t_3}) - {{\bf{Y}}^{[1]}}({t_2})}\\
 \vdots \\
{{{\bf{Y}}^{[1]}}({t_{{T_{AB}}}}) - {{\bf{Y}}^{[1]}}({t_2})}
\end{array}} \right] = \underbrace {\left[ {\begin{array}{*{20}{c}}
{{{\bf{H}}^{[11]}}({t_1})}&{{{\bf{H}}^{[12]}}({t_1})}\\
{{{\bf{H}}^{[11]}}({t_3}){\bf{V}}_1^{[1]}({t_3})}&{{{\bf{H}}^{[12]}}({t_3}){\bf{V}}_1^{[2]}({t_3})}\\
 \vdots & \vdots \\
{{{\bf{H}}^{[11]}}({t_{{T_{AB}}}}){\bf{V}}_1^{[1]}({t_{{T_{AB}}}})}&{{{\bf{H}}^{[12]}}({t_{{T_{AB}}}}){\bf{V}}_1^{[2]}({t_{{T_{AB}}}})}
\end{array}} \right]}_{{{{\bf{\hat H}}}_2}}\left[ {\begin{array}{*{20}{c}}
{{{\bf{u}}^{[1]}}}\\
{{{\bf{u}}^{[2]}}}
\end{array}} \right].
\end{equation}

Note that the elements of the precoding matrices ${\bf{V}}_1^{[1]}(n)$ and ${\bf{V}}_1^{[2]}(n)$  are generated from independent channel coefficients of ${{\bf{H}}^{[21]}}(n)$  and  ${{\bf{H}}^{[22]}}(n)$, respectively. And the channel matrices of the same path in different time slots belong to disparate channel blocks. Therefore, the effective channel matrix of ${{\bf{\hat H}}_2}$  has a full rank almost surely, i.e., rank(${{\bf{\hat H}}_2}$)=$2A$. Thus, receiver 1 is able to decode the $2A$  desired symbols $\{ u_1^{[1]},...,u_A^{[1]},u_1^{[2]},...,u_A^{[2]}\}$  by the end of phase two. Simultaneously, receiver 2 successfully decodes the  $2A$ desired symbols $\{ v_1^{[1]},...,v_A^{[1]},v_1^{[2]},...,v_A^{[2]}\}$. As a consequence, $4A$ sum-DoF is achievable over $\left| {{I_l}} \right| = {T_{AB}}$ channel uses.

Recall that a total number of time resources is ${S_t} = \{ 1,2,...,{T_{AB}}(n + {T_{AB}} - 1)\}$  and we have shown that $4An$ sum-DoF are achievable over $n$ time slot sets, i.e., $\left| {{I_1} \cup  \cdots  \cup {I_n}} \right| = {T_{AB}}n$. With the TDMA transmission method, we can achieve additional $\min (2A,B){T_{AB}}({T_{AB}} - 1)$ sum-DoF for the residual ${T_{AB}}({T_{AB}} - 1)$  time slots. Hence, we have 
\begin{equation}
d_\Sigma ^{{X_L}}(2,2;\frac{2}{{{T_{AB}}}}) = \frac{{4An + \min (2A,B){T_{AB}}({T_{AB}} - 1)}}{{{T_{AB}}n + {T_{AB}}({T_{AB}} - 1)}}.
\end{equation}
Therefore, as $n$ goes to infinity, the sum-DoF gain asymptotically achieves $\frac{{4A}}{{{T_{AB}}}}.$

b)$A<B<{2A}$.

For this case, we have $\frac{4A}{3}$  sum-DoF in total. The transmission scheme here is a little different from the proposed scheme for the case $B \le A$  because the precoding matrix constructed by \textit{Cyclic Zero-padding} under the condition $A<B<2A$ is unavailable. However, such a MIMO X channel is equivalent to a MIMO X channel with $A$ antennas at each node by switching off $B-A$ antennas at each receiver. Therefore, the coding scheme is straightforward by using a scaled version of the proposed scheme in [25]. It is remarkable that even in this setting the achievable sum-DoF can still be represented as the form in Theorem 1, i.e., $4A$ fresh symbols can be decoded successfully over 3 time slots and a total of $d_\Sigma ^{{X_L}}(2,2;\frac{2}{3}) = \frac{{4A}}{3}$  can be achievable.

c)$B\ge {2A}$

As the number of antennas at receivers becomes large, one can easily prove that the sum-DoF of $\frac{4A}{2}$  is achievable for any normalized CSI feedback delay because no more linear independent equations in desired symbols are required for each receiver, i.e., each receiver is able to decode $2A$ symbols over one channel use. Note that for this case, time sharing between the proposed scheme and a TDMA method is needless. Thus, the $a(A,B)\lambda  + b(A,B)$ in Theorem 1 will make no sense and the achievable trade-off region between the sum-DoF and $\lambda$ will be a straight line.

\textit{Remark 3 (CSI Feedback Delay):} We take $\frac{2}{T_{AB}}$  as an allowable normalized CSI feedback delay where the proposed transmission for the two-user MIMO X channel can achieve $\frac{4A}{T_{AB}}$  sum-DoF. In fact, the threshold for CSI feedback delay can be considered as an optimization problem with some constraints in detail. For example, in case  $B\le A$, an appropriate CSI feedback delay should be selected to ensure that $n$ enough time slot sets, i.e., $\{ {I_1},{I_2},...,{I_n}\}$, can be picked from the $n+T_{AB}-1$  channel blocks, where each time slot set $I_l$ has $T_{AB}$  elements for applying the proposed method. Nevertheless, the maximum allowable feedback delay achieving the optimal sum-DoF remains an open problem.

\textit{Remark 4 (An Extension to the $M\times2$  MIMO X Channel with a Symmetric Antenna Configuration):} A similar transmission scheme comprised of two phases can be easily proposed for the $M\times2$  MIMO X channel with a symmetric antenna configuration. Actually a total of $2MA$  independent symbols can be successfully decoded at each receiver over ${T_{AB}} = 2 + \left\lceil {\frac{{MA - B}}{B}} \right\rceil$  time slots where a certain allowable normalized CSI feedback delay $\lambda_{AB}$  is supposed. We claim that the sum-DoF $d_\Sigma ^{{X_L}}(M,2;{\lambda _{AB}}) = \frac{{2MA}}{{{T_{AB}}}}$  is achievable with local CSIT as long as the CSI feedback delay is less than  $\lambda_{AB}$. 

\textit{Remark 5 (An Extension to the Two-user MIMO X Channel with an Asymmetric Antenna Configuration):} We further discuss the two-user MIMO X channel in a more general setting where transmitter $i$ has $A_i$  antennas and receiver $j$ has $B_j$  antennas, for  $i,j\in\{1,2\}$. For case $\min\{A_i\} \ge \max\{B_i\}$, we argue that $A_1+A_2$  desired symbols will be resolved at each receiver over ${T_{{A_1}{A_2}{B_1}{B_2}}} = 2 + \left\lceil {\frac{{{A_1} + {A_2} - \min ({B_1},{B_2})}}{{\min ({B_1},{B_2})}}} \right\rceil$  time slots and $d_\Sigma ^{{X_L}}(2,2;{\lambda _{{A_1}{A_2}{B_1}{B_2}}}) = \frac{{2({A_1} + {A_2})}}{{{T_{{A_1}{A_2}{B_1}{B_2}}}}}$  is achievable as long as the CSI feedback delay is less than  $\lambda_{A_1A_2B_1B_2}$. The sum-DoF will be less than satisfactory when the gap between the numbers of antennas at each receiver is too large. This is because how many time slots required in phase two is determined by the receiver who has fewer antennas. 

\section{Transmission Scheme Achieving the Sum-DoF in Theorem 2}
In this section, we describe the transmission scheme that can achieve the sum-DoF in Theorem 2. To explain the basic idea, we elaborate the transmission scheme for the case of $K =3$. The discussion can be generalized to the general $K$-user scenario straightforwardly.  

For the three-user SISO X network, we will show that even with distributed and temperately delayed CSIT, $\frac{5}{4}$  sum-DoF is achievable with  $\lambda=\frac{2}{8}$. More precisely, a total of 15 independent information symbols will be successfully decoded at receivers during 12 channel uses. We select the time slots for the proposed transmission method in a same manner as section IV, i.e.,  $n \in \{ {t_1},{t_2},...,{t_{12}}\}$, where each time slot belongs to a different channel block, while $\{ {t_1},{t_4},{t_7}\}$  represent time slots without current CSIT and $\{ {t_2},{t_3},{t_5},{t_6},{t_8},{t_9}\}$  represent time slots when the transmitters have access to both current and delayed CSIT. We refer to $u, v, w$ as variables intended for receiver 1, 2, and 3, respectively. The proposed transmission scheme involves four phases.

\textit{Phase one:} Phase one is dedicated to receiver 1 and it spans three time slots, i.e.,   $n\in\{t_1,t_2,t_3\}$. In time slot  $t_1$, each transmitter $i\in\{1,2,3\}$ feeds a fresh information symbol $u_1^{[i]}$  to the channel. Therefore, at receiver  $j$, for  $j\in\{1,2,3\}$, we have
\begin{equation}
{y^{[j]}}({t_1}) = {h^{[j1]}}({t_1})u_1^{[1]} + {h^{[j2]}}({t_1})u_1^{[2]} + {h^{[j3]}}({t_1})u_1^{[3]}.
\end{equation}

In time slot  $t_2$, transmitter $i$  is able to exploit both current and outdated CSIT. Transmitter 1 sends another fresh information symbol $u_2^{[1]}$  for receiver 1, while transmitter 2 and 3 respectively construct the transmit signals as
\begin{equation}
{x^{[2]}}({t_2}) = \frac{{{h^{[22]}}({t_1})}}{{{h^{[22]}}({t_2} - 2)}}u_1^{[2]},{x^{[3]}}({t_2}) = \frac{{{h^{[23]}}({t_1})}}{{{h^{[23]}}({t_2} - 2)}}u_1^{[3]}.
\end{equation}
Since  ${{h^{[ji]}}({t_2})}={{h^{[ji]}}({t_2} - 2)}$, for  $i,j\in\{1,2,3\}$, the received signals can be written as
\begin{equation}
{y^{[j]}}({t_2}) = {h^{[j1]}}({t_2})u_2^{[1]} + {h^{[j2]}}({t_2})\frac{{{h^{[22]}}({t_1})}}{{{h^{[22]}}({t_2})}}u_1^{[2]}+ {h^{[j3]}}({t_2})\frac{{{h^{[23]}}({t_1})}}{{{h^{[23]}}({t_2})}}u_1^{[3]}.
\end{equation}

In time slot  $t_3$, similar operation is repeated. Transmitter 1 sends another fresh information symbol $u_3^{[1]}$  for receiver 1. Transmitter 2 and 3 simultaneously retransmit their previous symbols with a special precoding technique as
\begin{equation}
{x^{[2]}}({t_3}) = \frac{{{h^{[32]}}({t_1})}}{{{h^{[32]}}({t_3} - 2)}}u_1^{[2]},{x^{[3]}}({t_3}) = \frac{{{h^{[33]}}({t_1})}}{{{h^{[33]}}({t_3} - 2)}}u_1^{[3]}.
\end{equation}
Thus, at receiver  $j$, for $j\in\{1,2,3\}$  we have
\begin{equation}
{y^{[j]}}({t_3}) = {h^{[j1]}}({t_3})u_3^{[1]} + {h^{[j2]}}({t_3})\frac{{{h^{[32]}}({t_1})}}{{{h^{[32]}}({t_3})}}u_1^{[2]} + {h^{[j3]}}({t_3})\frac{{{h^{[33]}}({t_1})}}{{{h^{[33]}}({t_3})}}u_1^{[3]}.
\end{equation}

The main idea for designing the precoding coefficients is to allow the unintended receivers 2 and 3 to separately eliminate the variables  $u_1^{[2]}$ and $u_1^{[3]}$, thereby obtaining a linear combination in variables originated from the corresponding transmitter 1 for themselves. In particular, let $y^{[2]}(t_2)$  subtract $y^{[2]}(t_1)$  for receiver 2 to obtain a linear combination of $u_1^{[1]}$  and  $u_2^{[1]}$, i.e., ${{\hat y}^{[21]}} = {h^{[21]}}({t_2})u_2^{[1]} - {h^{[21]}}({t_1})u_1^{[1]}$. Likewise, receiver 3 obtains a new combination ${\hat y^{[31]}}$  comprising of $u_1^{[1]}$ and $u_3^{[1]}$  by subtracting $y^{[3]}(t_1)$  from  $y^{[3]}(t_3)$, i.e., ${{\hat y}^{[31]}} = {h^{[31]}}({t_3})u_3^{[1]} - {h^{[31]}}({t_1})u_1^{[1]}$. Note that ${\hat y^{[21]}}$  and ${\hat y^{[31]}}$  are linearly independent almost surely, each of which can be reconstructed by transmitter 1 with local CSIT.

\textit{Phase two:} Phase two is dedicated to receiver 2 and it also spans three time slots, i.e.,  $n \in \{ {t_4},{t_5},{t_6}\}$. In each time slot during phase two, transmitter 2 feeds a fresh information symbol  $v_1^{[2]}$, $v_2^{[2]}$  and  $v_3^{[2]}$, respectively. Transmitter 1 sends an information symbol $v_1^{[1]}$  for receiver 2 in time slot $t_4$  and retransmit it after precoding in the next two time slots $t_5$  and  $t_6$. Transmitter 3 does the similar operation as transmitter 1. The transmit signals can be described as
\begin{eqnarray}
{x^{[1]}}({t_5}) = \frac{{{h^{[11]}}({t_4})}}{{{h^{[11]}}({t_5} - 2)}}v_1^{[1]},{x^{[3]}}({t_5}) = \frac{{{h^{[13]}}({t_4})}}{{{h^{[13]}}({t_5} - 2)}}v_1^{[3]},\nonumber \\
{x^{[1]}}({t_6}) = \frac{{{h^{[31]}}({t_4})}}{{{h^{[31]}}({t_6} - 2)}}v_1^{[1]},{x^{[3]}}({t_6}) = \frac{{{h^{[33]}}({t_4})}}{{{h^{[33]}}({t_6} - 2)}}v_1^{[3]}.
\end{eqnarray}

Therefore, at receiver $j$, for $j\in\{1,2,3\}$, we have
\begin{eqnarray}
{y^{[j]}}({t_4}) &=& {h^{[j1]}}({t_4})v_1^{[1]} + {h^{[j2]}}({t_4})v_1^{[2]} + {h^{[j3]}}({t_4})v_1^{[3]}, \nonumber \\
{y^{[j]}}({t_5}) &=& {h^{[j1]}}({t_5}){x^{[1]}}({t_5}) + {h^{[j2]}}({t_5})v_2^{[2]} + {h^{[j3]}}({t_5}){x^{[3]}}({t_5}),\nonumber \\
{y^{[j]}}({t_6}) &=& {h^{[j1]}}({t_6}){x^{[1]}}({t_6}) + {h^{[j2]}}({t_6})v_3^{[2]} + {h^{[j3]}}({t_6}){x^{[3]}}({t_6}).
\end{eqnarray}
Since ${h^{[ji]}}({t_5}) = {h^{[ji]}}({t_5} - 2),{h^{[ji]}}({t_6}) = {h^{[ji]}}({t_6} - 2)$  for  $i,j\in\{1,2,3\}$, receiver 1 and 3 can do the similar operation as phase one to obtain a linear combination of variables originating from transmitter 2, respectively. In particular, let $y^{[1]}(t_5)$  subtract $y^{[1]}(t_4)$  for receiver 1 to obtain a linear combination, i.e., ${{\hat y}^{[12]}} = {h^{[12]}}({t_5})v_2^{[2]} - {h^{[12]}}({t_4})v_1^{[2]}$. Likewise, by subtracting  $y^{[3]}(t_4)$ from  $y^{[3]}(t_6)$, receiver 3 can obtain a new combination, i.e., ${{\hat y}^{[32]}} = {h^{[32]}}({t_6})v_3^{[2]} - {h^{[32]}}({t_4})v_1^{[2]}$. Note that ${\hat y^{[12]}}$  and ${\hat y^{[32]}}$  are linearly independent almost surely, each of which can be reconstructed by transmitter 2 with local CSIT.

\textit{Phase three:} Phase three is dedicated to receiver 3 and it also spans three time slots, i.e.,  $n\in\{t_7,t_8,t_9\}$. Similar to phase one and two, in each time slot, transmitter 3 feeds a fresh information symbol $w_1^{[3]},w_2^{[3]},w_3^{[3]}$  for receiver 3 respectively. Transmitter 1 sends an information symbol $w_1^{[1]}$  for receiver 3 in its first time slot and retransmit it after precoding in each of the next two time slots. Transmitter 2 does the similar operation as transmitter 1. Therefore, the transmitted and received signals can be described as
\begin{eqnarray}
\!\!\!\!\!\!\!{x^{[1]}}({t_8}) &=& \frac{{{h^{[11]}}({t_7})}}{{{h^{[11]}}({t_8} - 2)}}w_1^{[1]},{x^{[2]}}({t_8}) = \frac{{{h^{[12]}}({t_7})}}{{{h^{[12]}}({t_8} - 2)}}w_1^{[2]},\nonumber \\
\!\!\!\!\!\!\!{x^{[1]}}({t_9}) &=& \frac{{{h^{[21]}}({t_7})}}{{{h^{[21]}}({t_9} - 2)}}w_1^{[1]},{x^{[2]}}({t_9}) = \frac{{{h^{[22]}}({t_7})}}{{{h^{[22]}}({t_9} - 2)}}w_1^{[2]}.
\end{eqnarray}

Thus, at receiver  $j$, for  $j\in\{1,2,3\}$, we have
\begin{eqnarray}
{y^{[j]}}({t_7}) &=& {h^{[j1]}}({t_7})w_1^{[1]} + {h^{[j2]}}({t_7})w_1^{[2]} + {h^{[j3]}}({t_7})w_1^{[3]},\nonumber \\
{y^{[j]}}({t_8}) &=& {h^{[j1]}}({t_8}){x^{[1]}}({t_8}) + {h^{[j2]}}({t_8}){x^{[2]}}({t_8}) + {h^{[j3]}}({t_8})w_2^{[3]}, \nonumber \\
{y^{[j]}}({t_9}) &=& {h^{[j1]}}({t_9}){x^{[1]}}({t_9}) + {h^{[j2]}}({t_9}){x^{[2]}}({t_9}) + {h^{[j3]}}({t_9})w_3^{[3]}.
\end{eqnarray}
Using the fact that ${h^{[ji]}}({t_8}) = {h^{[ji]}}({t_8} - 2),{h^{[ji]}}({t_9}) = {h^{[ji]}}({t_9} - 2)$  for $i,j\in \{1,2,3\}$, receiver 1 and 2 can do the similar operation as phase one and two to obtain a linear combination of variables originating from transmitter 3. In particular, let $y^{[1]}(t_8)$  subtract $y^{[1]}(t_7)$  for receiver 1 to obtain a linear combination, i.e., ${{\hat y}^{[13]}} = {h^{[13]}}({t_8})w_2^{[3]} - {h^{[13]}}({t_7})w_1^{[3]}$. Likewise, receiver 2 can do the similar operation via $y^{[2]}(t_9)$  minus  $y^{[2]}(t_7)$ to get a new combination, i.e., ${{\hat y}^{[23]}} = {h^{[23]}}({t_9})w_3^{[3]} - {h^{[23]}}({t_7})w_1^{[3]}$. Note that ${\hat y^{[13]}}$  and ${\hat y^{[23]}}$  are linearly independent almost surely, each of which can be reconstructed by transmitter 3 with local CSIT.

\textit{Phase four:} Phase four consists of three time slots, i.e., $n\in\{t_{10},t_{11},t_{12}\}$. Recall that each transmitter by the end of phase three is able to reconstruct the corresponding new combinations. In time slot  $t_{10}$, transmitter 1 transmits  ${\hat y^{[21]}}$, transmitter 2 transmits  ${\hat y^{[12]}}$, and transmitter 3 keeps silent. In time slot  $t_{11}$, ${\hat y^{[31]}}$   and ${\hat y^{[13]}}$   are sent from transmitters 1 and 3 respectively. In time slot  $t_{12}$, transmitter 2 sends ${\hat y^{[32]}}$   and transmitter 3 sends  ${\hat y^{[23]}}$. 

Consider receiver 1. From the linear combinations of  ${\hat y^{[21]}}$  and  ${\hat y^{[12]}}$  received over the first time slot in phase four, it is able to remove  ${\hat y^{[12]}}$  that previously acquired in phase two, to obtain   ${\hat y^{[21]}}$. Likewise, receiver 1 has access to   ${\hat y^{[31]}}$. Thus, receiver 1 has
\begin{equation}
\left[ {\begin{array}{*{20}{c}}
{\begin{array}{*{20}{c}}
{{y^{[1]}}({t_1})}\\
{{y^{[1]}}({t_2})}
\end{array}}\\
{{y^{[1]}}({t_3})}\\
{{{\hat y}^{[21]}}}\\
{{{\hat y}^{[31]}}}
\end{array}} \right] = \underbrace {\left[ {\begin{array}{*{20}{c}}
{{h^{[11]}}({t_1})}&0&0&{{h^{[12]}}({t_1})}&{{h^{[13]}}({t_1})}\\
0&{{h^{[11]}}({t_2})}&0&{{h^{[12]}}({t_2})\frac{{{h^{[22]}}({t_1})}}{{{h^{[22]}}({t_2})}}}&{{h^{[13]}}({t_2})\frac{{{h^{[23]}}({t_1})}}{{{h^{[23]}}({t_2})}}}\\
0&0&{{h^{[11]}}({t_3})}&{{h^{[12]}}({t_3})\frac{{{h^{[32]}}({t_1})}}{{{h^{[32]}}({t_3})}}}&{{h^{[13]}}({t_3})\frac{{{h^{[33]}}({t_1})}}{{{h^{[33]}}({t_3})}}}\\
{ - {h^{[21]}}({t_1})}&{{h^{[21]}}({t_2})}&0&0&0\\
{ - {h^{[31]}}({t_1})}&0&{{h^{[31]}}({t_3})}&0&0
\end{array}} \right]}_{{{{\bf{\hat H}}}_3}}\left[ {\begin{array}{*{20}{c}}
{\begin{array}{*{20}{c}}
{u_1^{[1]}}\\
{u_2^{[1]}}
\end{array}}\\
{u_3^{[1]}}\\
{u_1^{[2]}}\\
{u_1^{[3]}}
\end{array}} \right].
\end{equation}

Since the channel coefficients are picked from a continuous random distribution and each time slot of $\{t_1,t_2,t_3\}$  belongs to a different channel block, the efficient channel matrix ${{\bf{\hat H}}_3}$  has a full rank almost surely, i.e., rank(${{\bf{\hat H}}_3}$)=5. Thus, receiver 1 can successfully decode these five variables. In a similar way, receiver 2 and receiver 3 can resolve the five variables $\{ v_1^{[1]},v_1^{[2]},v_2^{[2]},v_3^{[2]},v_1^{[3]}\}$  and $\{ w_1^{[1]},w_1^{[2]},w_1^{[3]},w_2^{[3]},w_3^{[3]}\}$, respectively. Thus, 15 transmitted information symbols are resolved over 12 channel uses and $\frac{5}{4}$  sum-DoF is achieved on the three-user SISO X channel. 

\textit{Remark 6 (Time Slots for the Transmission Scheme):} Note that each phase is mutually independent in channel path as well as the transmitted symbols. For instance, ${\hat y^{[21]}}$  is in the form of  ${h^{[21]}}({t_2})u_2^{[1]} - {h^{[21]}}({t_1})u_1^{[1]}$, while ${\hat y^{[12]}}$ is consisted of channel coefficients in different path and symbols dedicated to receiver 2, i.e., ${h^{[12]}}({t_5})v_2^{[2]} - {h^{[12]}}({t_4})v_1^{[2]}$. Thus, there is no need to select a time slot set as each time slot in this set belongs to a different block. What we need to ensure is that the time slots for each of the previous three phases are picked from different channel blocks, such as $\{t_1,t_2,t_3\}$  for phase one should be selected from three different channel blocks. Moreover, the time slots $\{t_{10},t_{11},t_{12}\}$ for the final phase can be selected from  arbitrary channel blocks because the final phase aiming to deliver the previous auxiliary linear combinations needs no any CSIT.  

\textit{Remark 7 (Overhead of channel feedback):} In each of the first three phases, only four complex values representing the ratios of the CSI need to be fed back. For instance, in phase one, only required information at transmitters in time slot $t_2$  and $t_3$  are effective channel values for precoding, i.e., $\left\{ {\frac{{{h^{[22]}}({t_1})}}{{{h^{[22]}}({t_2} - 2)}},\frac{{{h^{[32]}}({t_1})}}{{{h^{[32]}}({t_3} - 2)}}} \right\}$  for transmitter 2 and $\left\{ {\frac{{{h^{[23]}}({t_1})}}{{{h^{[23]}}({t_2} - 2)}},\frac{{{h^{[33]}}({t_1})}}{{{h^{[33]}}({t_3} - 2)}}} \right\}$  for transmitter 3. Thus, a more practical precoding technique with reduced CSI feedback amount is proposed to achieve the same sum-DoF for the three-user SISO X network compared to [11]. This implies that global and delayed CSIT is not necessarily required to obtain the greater sum-DoF than that achievable with no CSIT.

\textit{Remark 8 (An Extension to the $K$-user SISO X Network):} The novelty of RIA over the $K$-user SISO X network appears in the construction of auxiliary linear combinations of independent information symbols that aid in the decoding of the previously transmitted information symbols based on only the information symbols and local CSIT available to each transmitter. With the idea used for the proof of the 3-user SISO X network, one can easily prove that the sum-DoF of $\frac{{2(2K - 1)}}{{3K - 1}}$ is achievable for the $K$-user SISO X network almost surely when the transmitters have local CSI
and the normalized feedback delay is less than $\frac{2}{{3K - 1}}$.

\section{Transmission Scheme Achieving the Sum-DoF in Theorem 3}
In this section, we explore the achievability of sum-DoF for the \textit{M}$\times$\textit{N} user MISO X network where each transmitter has $A=N-1$  antennas and each receiver has a single antenna. 

We focus on the proof of the point  $d_\Sigma ^{{X_L}}(M,N;\frac{2}{{M(N - 1) + 1}}) = \frac{{MN(N - 1)}}{{M(N - 1) + 1}}$, where the feedback delay is $T_{fb}=2$  and the channel coherence time becomes  $T_c=M(N-1)+1$. Over the $M(N-1)+1$  channel uses, the proposed scheme achieves $N-1$  degrees of freedom for each of the $MN$  messages  $W^{[ji]}$,  $i\in\{1,2,...,M\}$,  $j\in\{1,2,...,N\}$. To show this, we consider $n+T_c-1$  channel blocks consisting of $(n+T_c-1)T_c$  time slots where we divide the time resources into two sets, $S_d$  with $\left| {{S_d}} \right| = 2(n + {T_c} - 1)$  and $S_c$  with $\left| {{S_c}} \right| = ({T_c} - 2)(n + {T_c} - 1)$. We further define $n$ time slot sets, $\{ {I_1},{I_2},...,{I_n}\}$, each of which has $T_c$  elements for applying the proposed method, i.e., ${I_l} = \{ {t_{l,1}},{t_{l,2}},...,{t_{l,{T_c}}}\}$  where $l \in \{ 1,2,...,n\}$,  $\{ {t_{l,1}}\}  \in {S_d}$, and ${t_{l,k}} \in {S_c}$  for $k \in \{ 2,3,...,{T_c}\}$. Remember that any two time slots of $I_l$  belong to difference channel blocks. Here we omit the index $l$ for simplicity, i.e., ${I_l} = \{ {t_1},{t_2},...,{t_{{T_c}}}\}$. The achievable scheme is as follows:

\textit{Phase one:} This phase takes one time slot, i.e.,  $n\in\{t_1\}$. Each transmitter sends a superposition of $A$-symbol vectors dedicated to all the receivers. We denote the transmitted signal as

\begin{equation}
{{\bf{X}}^{[i]}}({t_1}) = \sum\limits_{j = 1}^N {{{\bf{s}}^{[ji]}}},
\end{equation}
where ${{\bf{s}}^{[ji]}} = {[s_1^{[ji]},s_2^{[ji]},...,s_A^{[ji]}]^T}$  is the signal vector from transmitter $i$ to receiver $j$, for  $i \in \{ 1,2,...,M\} ,j \in \{ 1,2,...,N\}$. The received signal at receiver $j$ will be
\begin{eqnarray}
\!\!\!\!\!\!{y^{[j]}}({t_1}) &=& \sum\limits_{i = 1}^M {{{\bf{h}}^{[ji]}}({t_1}){{\bf{X}}^{[i]}}({t_1})} ,\nonumber \\
&=& \underbrace {\sum\limits_{i = 1}^M {{{\bf{h}}^{[ji]}}({t_1}){{\bf{s}}^{[ji]}}} }_{desired} + \underbrace {\sum\limits_{i = 1}^M {{{\bf{h}}^{[ji]}}({t_1})\left( {\sum\limits_{k = 1,k \ne j}^N {{{\bf{s}}^{[ki]}}} } \right)} }_{undesired}.
\end{eqnarray}
where  $j\in\{1,2,...,N\}$, and ${{\bf{h}}^{[ji]}}({t_1}) = [h_1^{[ji]}({t_1}),...,h_A^{[ji]}({t_1})]$  is a $1\times{A}$  vector representing the channel vector from transmitter $i$ to receiver $j$ in time slot  $t_1$. By the end of phase one, each receiver obtains a linear equation involving two items, i.e., desired terms and undesired (interfering) terms.

\textit{Phase two:} Here comes the preparatory phase for interference cancellation at receivers. The superposition of $A$-symbol vectors is retransmitted in each time slot $n$, for $n \in \{ {t_2},...,{t_{{T_c}}}\}$, with precoding matrices. In other words, during a time slot $n$, message $W^{[ji]}$  is encoded at transmitter $i$ as $A$ independent streams $s_a^{[ji]}$  along directions ${\bf{v}}_a^{[ji]}(n)$, for  $a=1,2,...,A$. So the signal transmitted at transmitter $i$  may be written as 
\begin{equation}
{{\bf{X}}^{[i]}}(n) = \sum\limits_{j = 1}^N {\sum\limits_{a = 1}^A {s_a^{[ji]}{\bf{v}}_a^{[ji]}(n)} }  = \sum\limits_{j = 1}^N {{\bf{V}}_j^{[i]}(n){{\bf{s}}^{[ji]}}},
\end{equation}
Note that ${\bf{V}}_j^{[i]}(n)$  is a $A\times{A}$  matrix whose columns are  ${\bf{v}}_a^{[ji]}(n)$,   $a=1,2,...,A$. In time slot $n$, the received signal at receiver  $j$,  $j\in\{1,2,...,N\}$, can then be written as 
\begin{eqnarray}
{y^{[j]}}(n) &=& \sum\limits_{i = 1}^M {{{\bf{h}}^{[ji]}}(n)\left( {\sum\limits_{j = 1}^N {{\bf{V}}_j^{[i]}(n){{\bf{s}}^{[ji]}}} } \right)} ,\nonumber \\
 &=& \underbrace {\sum\limits_{i = 1}^M {{{\bf{h}}^{[ji]}}(n){\bf{V}}_j^{[i]}(n){{\bf{s}}^{[ji]}}} }_{desired} + \underbrace {\sum\limits_{i = 1}^M {{{\bf{h}}^{[ji]}}(n)\left( {\sum\limits_{k = 1,k \ne j}^N {{\bf{V}}_k^{[i]}(n){{\bf{s}}^{[ki]}}} } \right)} }_{undesired}.
\end{eqnarray}

We wish to design precoding matrices ${\bf{V}}_k^{[i]}(n)$  so that receiver $j$ can eliminate the undesired item by interference cancellation. Once the interference is eliminated by subtracting ${y^{[j]}}({t_1})$ from  ${y^{[j]}}(n)$, a receiver can obtain a linear equation in $MA$ desired symbols. Repeating the same operation for the residual time slots during phase two, there will be $M(N-1)$  linear equations in $MA$ desired symbols observed at receiver $j$. Interference cancellation is ensured by constructing the precoding matrices ${\bf{V}}_k^{[i]}(n)$  so that the following conditions (on the left) are satisfied at receiver $j$, for  $j\in\{1,2,...,N\}$:
\begin{equation}
\begin{array}{l}
\left. \begin{array}{c}
{{\bf{h}}^{[ji]}}(n){\bf{V}}_1^{[i]}(n) = {{\bf{h}}^{[ji]}}({t_1})\\
{{\bf{h}}^{[ji]}}(n){\bf{V}}_2^{[i]}(n) = {{\bf{h}}^{[ji]}}({t_1})\\
 \vdots \\
{{\bf{h}}^{[ji]}}(n){\bf{V}}_{j - 1}^{[i]}(n) = {{\bf{h}}^{[ji]}}({t_1})\\
{{\bf{h}}^{[ji]}}(n){\bf{V}}_{j + 1}^{[i]}(n) = {{\bf{h}}^{[ji]}}({t_1})\\
 \vdots \\
{{\bf{h}}^{[ji]}}(n){\bf{V}}_N^{[i]}(n) = {{\bf{h}}^{[ji]}}({t_1})
\end{array} \right\} \Leftrightarrow \left\{ \begin{array}{c}
{{\bf{h}}^{[1i]}}(n){\bf{V}}_k^{[i]}(n) = {{\bf{h}}^{[1i]}}({t_1})\\
{{\bf{h}}^{[2i]}}(n){\bf{V}}_k^{[i]}(n) = {{\bf{h}}^{[2i]}}({t_1})\\
 \vdots \\
{{\bf{h}}^{[(k - 1)i]}}(n){\bf{V}}_k^{[i]}(n) = {{\bf{h}}^{[(k - 1)i]}}({t_1})\\
{{\bf{h}}^{[(k + 1)i]}}(n){\bf{V}}_k^{[i]}(n) = {{\bf{h}}^{[(k + 1)i]}}({t_1})\\
 \vdots \\
{{\bf{h}}^{[Ni]}}(n){\bf{V}}_k^{[i]}(n) = {{\bf{h}}^{[Ni]}}({t_1})
\end{array} \right.\\
\begin{array}{*{20}{c}}
{\begin{array}{*{20}{c}}
{\quad\quad\forall i \in \{ 1,2,...,M\} .}&{\quad\quad\quad\quad}
\end{array}}&{\begin{array}{*{20}{c}}
{\begin{array}{*{20}{c}}
{}&{}
\end{array}}&{\forall i \in \{ 1,2,...,M\} ,\forall k \in \{ 1,2,...,N\} .}
\end{array}}
\end{array}
\end{array}
\end{equation}
In other words, we wish to construct precoding matrices ${\bf{V}}_k^{[i]}(n)$  so that, at receiver $j$, all the effective channel vectors ${{\bf{h}}^{[ji]}}(n){\bf{V}}_k^{[i]}(n)$  carrying the interference originated from transmitters $i\in\{1,2,...,M\}$  in time slot $n$ can be equal to the channel vectors ${{\bf{h}}^{[ji]}}({t_1})$  previously seen in time slot  $t_1$. Note that there are $A=N-1$  relations above for a certain  ${\bf{V}}_k^{[i]}(n)$. These relations can be recorded to be expressed alternately as the right side above. Remember that ${{\bf{h}}^{[ji]}}(n)$  are $1\times{A}$  vectors, the relations can be rewritten as 
\begin{equation}
{\bf{V}}_k^{[i]}(n) = \underbrace {{{\left[ {\begin{array}{*{20}{c}}
{\begin{array}{*{20}{c}}
{{{\bf{h}}^{[1i]}}(n)}\\
 \vdots 
\end{array}}\\
{\begin{array}{*{20}{c}}
{{{\bf{h}}^{[(k - 1)i]}}(n)}\\
{{{\bf{h}}^{[(k + 1)i]}}(n)}
\end{array}}\\
 \vdots \\
{{{\bf{h}}^{[Ni]}}(n)}
\end{array}} \right]}^{ - 1}}}_{{\bf{\hat H}}(n)}\underbrace {\left[ {\begin{array}{*{20}{c}}
{\begin{array}{*{20}{c}}
{\begin{array}{*{20}{c}}
{\begin{array}{*{20}{c}}
{\begin{array}{*{20}{c}}
{{{\bf{h}}^{[1i]}}({t_1})}\\
 \vdots 
\end{array}}\\
{{{\bf{h}}^{[(k - 1)i]}}({t_1})}
\end{array}}\\
{{{\bf{h}}^{[(k + 1)i]}}({t_1})}
\end{array}}\\
 \vdots 
\end{array}}\\
{{{\bf{h}}^{[Ni]}}({t_1})}
\end{array}} \right]}_{{\bf{\hat H}}({t_1})}.
\end{equation}

Now the same interference pattern at receiver $j$ before and after phase two is guaranteed because ${\bf{\hat H}}(n) \in {\mathbb{C}^{A \times A}}$  has a full rank almost surely. Thus, each receiver $j\in\{1,2,...,N\}$ is able to extract a desired equation by subtracting  ${y^{[j]}}({t_1})$ from ${y^{[j]}}(n)$, i.e.,
\begin{equation}
{y^{[j]}}(n) - {y^{[j]}}({t_1}) = \sum\limits_{i = 1}^M {{{\bf{h}}^{[ji]}}(n){\bf{V}}_j^{[i]}(n){{\bf{s}}^{[ji]}}}  - \sum\limits_{i = 1}^M {{{\bf{h}}^{[ji]}}({t_1}){{\bf{s}}^{[ji]}}}.
\end{equation}
Finally, by the end of phase two, a receiver can obtain $MA$ linear equations of $MA$ variables. We describe the effective channel input-output relationship for receiver $j$ during the selected time slot set $I_l$ as
\begin{equation}
\left[ {\begin{array}{*{20}{c}}
{{y^{[j]}}({t_2}) \!-\! {y^{[j]}}({t_1})}\\
{{y^{[j]}}({t_3}) \!-\! {y^{[j]}}({t_1})}\\
 \vdots \\
{{y^{[j]}}({t_{{T_c}}}) \!-\! {y^{[j]}}({t_1})}
\end{array}} \right] \!\!=\!\! \underbrace {\left[ {\begin{array}{*{20}{c}}
{\begin{array}{*{20}{c}}
{{{\bf{h}}^{[j1]}}({t_2}){\bf{V}}_j^{[1]}({t_2}) \!-\! {{\bf{h}}^{[j1]}}({t_1})}\\
{{{\bf{h}}^{[j1]}}({t_3}){\bf{V}}_j^{[1]}({t_3}) \!-\! {{\bf{h}}^{[j1]}}({t_1})}\\
 \vdots \\
{{{\bf{h}}^{[j1]}}({t_{{T_c}}}){\bf{V}}_j^{[1]}({t_{{T_c}}}) \!-\! {{\bf{h}}^{[j1]}}({t_1})}
\end{array}}&{\begin{array}{*{20}{c}}
 \cdots \\
 \cdots \\
 \ddots \\
 \cdots 
\end{array}}&{\begin{array}{*{20}{c}}
{{{\bf{h}}^{[jM]}}({t_2}){\bf{V}}_j^{[M]}({t_2}) \!-\! {{\bf{h}}^{[jM]}}({t_1})}\\
{{{\bf{h}}^{[jM]}}({t_3}){\bf{V}}_j^{[M]}({t_3}) \!-\! {{\bf{h}}^{[jM]}}({t_1})}\\
 \vdots \\
{{{\bf{h}}^{[jM]}}({t_{{T_c}}}){\bf{V}}_j^{[M]}({t_{{T_c}}}) \!-\! {{\bf{h}}^{[jM]}}({t_1})}
\end{array}}
\end{array}} \right]}_{{{{\bf{\hat H}}}_5}}\left[ {\begin{array}{*{20}{c}}
{{{\bf{s}}^{[j1]}}}\\
{{{\bf{s}}^{[j2]}}}\\
 \vdots \\
{{{\bf{s}}^{[jM]}}}
\end{array}} \right].
\end{equation}

Recall that the precoding matrices ${\bf{V}}_j^{[i]}(n)$  for $n\in \{ {t_2},...,{t_{{T_c}}}\}$  were generated independently from channel ${{\bf{h}}^{[ji]}}(n)$  and each time slot $n$ belongs to a different channel block. Further, the elements of the channel vectors are picked from a continuous random distribution. Therefore, the channel vectors ${{\bf{h}}^{[j1]}}({t_1}), \ldots ,{{\bf{h}}^{[jM]}}({t_1})$  and ${{\bf{h}}^{[j1]}}(n){\bf{V}}_j^{[1]}(n), \ldots ,{{\bf{h}}^{[jM]}}(n){\bf{V}}_j^{[M]}(n)$ for  $n \in \{ {t_2},...,{t_{{T_c}}}\}$, are statistically independent and the effective channel matrix ${{\bf{\hat H}}_5}$  has a full rank almost surely, i.e., rank(${{\bf{\hat H}}_5}$)=$MA$. Lastly, receiver $j$ successfully decodes $MA$ desired symbols over $\left| {{I_l}} \right| = {T_c}$  channel uses. For the other time resources we simply apply a TDMA transmission method. Thus, we have
\begin{equation}
d_\Sigma ^{{X_L}}(M,N;\frac{2}{{{T_c}}}) = \frac{{MN(N - 1)n + {T_c}({T_c} - 1)}}{{{T_c}n + {T_c}({T_c} - 1)}},
\end{equation}
where the asymptotical sum-DoF gain is $\frac{{MN(N - 1)}}{{M(N - 1) + 1}}$ as $n$ goes to infinity.

\textit{Remark 9 (An Inner Bound on the Sum-DoF of MISO X Networks):} We are able to establish an inner bound with the achievability proof of theorem 3, on the condition of temperately-delayed CSI feedback, thereby revealing insights that this bound is tight for the $A=N-1$  case, for which we achieve the bound of $\min (N,\frac{{MN(N - 1)}}{{M(N - 1) + 1}})$ in the \textit{M}$\times$\textit{N}  MISO X network with full CSI [4]. Although the bound does not scale with neither $M$ nor $N$, the number of transmitters or receivers, it's the best known inner bound under the local and temperately-delayed CSIT. 

\section{Conclusion}
In this paper, we investigated achievable sum-DoFs of multiuser MIMO X networks including the two-user MIMO X network, $K$-user SISO X network and \textit{M}$\times$\textit{N}  user MIMO X network with local and temperately-delayed CSIT. With the proposed precoding technique (\textit{Cyclic Zero-padding}), we are able to characterize the achievable trade-off between the sum-DoF and CSI feedback delay in the two-user MIMO X channel with distributed CSIT. We also showed that the RIA scheme can improve the performance of DoF with local and temperately-delayed CSIT in the $K$-user SISO X network. Finally, an achievable DoF inner bound of the \textit{M}$\times$\textit{N}  user MISO X network was developed. We note that this inner bound is tight when each transmitter has $N-1$ antennas and each receiver has a single antenna. Based on the spatial scale invariance property, we extended the results to the general MIMO scenario. It is interesting to observe that the multiuser MIMO X networks with various antenna configurations allow us to design efficient transmission schemes that can achieve larger sum-DoF by using less time slots. 

\appendices
\section{Cyclic Zero-Padding}
First, we would like to develop a lemma that will be useful in the description of \textit{Cyclic Zero-padding} which leads to the construction of precoding matrices for the $B\le A$  case in the two-user MIMO X channel. 

\textit{Lemma 1:} Consider an $A\times A$  square matrix $G$ such that  $g_{ij}$, the element in the $i$th row and $j$th column of $G$, is of the form as
\begin{equation}
{g_{ij}} = \left\{ {\begin{array}{*{20}{c}}
{{f_{ij}}({x^{[\alpha _{_1}^{[j]}]}},...,{x^{[\alpha _{_{BB}}^{[j]}]}},{y^{[\gamma _1^{[j]}]}},...,{y^{[\gamma _B^{[j]}]}}),}\\
{0,}
\end{array}} \right.\begin{array}{*{20}{c}}
{\begin{array}{*{20}{c}}
{}&{\!\!\!\!{\rm{for  }}\,\,\, i \in \{ 1,...,B - j + 1\}  \cup \{ A - j + 2,...,A\} ,}
\end{array}}\\
{\begin{array}{*{20}{c}}
{\begin{array}{*{20}{c}}
{}&{\!\!\!\!\!\!\!\!\!\!\!\!\!\!\!\!\!\!\!\!\!\!\!\!\!\!\!\!\!\!\!\!\!\!\!\! { \rm{for  }}\quad i \in \{ B - j + 2,...,A - j + 1\} \cup \{A+B-j+2,...,A\} }
\end{array}.}&{}
\end{array}}
\end{array}
\end{equation} 
where we define ${S_x} = \{ {x^{[1]}},...,{x^{[AB]}}\}$ and ${S_y} = \{ {y^{[1]}},...,{y^{[AB]}}\}$  as two sets of i.i.d. random variables drawn from a continuous random distribution. Further, we divide $S_x$  into $A$ subsets, $\{ S_x^{[1]},S_x^{[2]},...,S_x^{[A]}\}$, each of which has $B^2$  elements, i.e., $S_x^{[j]} = \{ {x^{[\alpha _{_1}^{[j]}]}},...,{x^{[\alpha _{_{BB}}^{[j]}]}}\}$  where  $j\in\{1,...,A\}$,  $\alpha _{_1}^{[j]},...,\alpha _{_{BB}}^{[j]} \in \{ 1,...,AB\}$. Likewise,  $S_y$ can be divided into $A$ subsets,  $\{ S_y^{[1]},S_y^{[2]},...,S_y^{[A]}\}$, each of which has $B$ elements, i.e., $S_y^{[j]} = \{ {y^{[\gamma _1^{[j]}]}},...,{y^{[\gamma _B^{[j]}]}}\}$  where  $j\in \{1,...,A\}$,  $\gamma _1^{[j]},...,\gamma _B^{[j]} \in \{ 1,...,AB\}$. Note that the elements of any two subsets from  $S_x^{[j]}$ overlap because  $\left| {{S_x}} \right| = AB$.  ${f_{ij}}(*)$ is an unique function corresponding to an element $g_{ij}$  and containing four algorithms of multiplication, division, addition and subtraction. In that way, if the absolute value of each element generated from the function  ${f_{ij}}(*)$ is bounded between a non-zero minimum value and a finite maximum value, the matrix $G$  has a full rank of $A$ with probability 1, of which the special type is shown in Fig. 6, where the subscripts $\alpha,\beta$  are some short-hands of different rows and columns.
\begin{figure}[]
\centering
\includegraphics[width=2in]{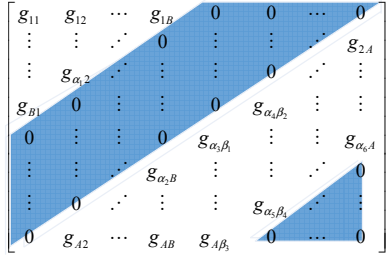}
\caption{Illustration of the matrix $G$.}
\end{figure}

\textit{Proof:} Starting from the perspective of determinant of a matrix, we need to show that det($G$) is nonzero with probability 1. We first discuss the constitution of this matrix where the elements in the blue zone are zeroes and each of the others is a nonzero random variable. As shown in Fig. 6, we see that the first column of the matrix is comprised of $B$ random variables above and $A-B$ zeroes below. When it comes to the second column, the number of variables and zeroes remain invariant excepting the relative positions, i.e., the zeroes in the second column shift upwards one unit as a whole comparing with those in the first column. Meanwhile, the overflowing element on the top fills the gap underneath in sequence. The following columns can be treated in a similar way until the zeroes move to the top position as a whole, i.e., after shifting $B$ times, there will be $B+1$ columns in total. Then, from the $(B+1)$th column, the zeroes continue to shift upwards for $A-B-2$ more times in the same manner resulting in $A$ columns in total finally. It is remarkable that each nonzero element in each column differs in the distribution since each of them is generated from a disparate function which is composed of a set of i.i.d. random variables by four algorithms. 

Let $D_{ij}$  denote the cofactor corresponding to $g_{ij}$  and removing the terms with zero coefficients. Then 
\begin{equation}
\det (G) = {D_{11}}{g_{11}} + {D_{12}}{g_{12}} + ... + {D_{1B}}{g_{1B}}.
\end{equation}
Recall that we assume each element generated from the function $f_{ij}(*)$  is nonzero with probability 1. Thus, det($G$)=0 only if a polynomial in such a set of i.i.d. random variables whose coefficients are  $D_{1j}$,  $j\in\{1,...,B\}$, is equal to zero. Therefore, det($G$)=0 with nonzero probability implies one of the following two events:

1) The i.i.d. random variables of $S_x^{[j]}$  and $S_y^{[j]}$  are roots of the polynomial formed by setting det($G$)=0.

2) The polynomial is the zero polynomial.\\
Note that the probability of these i.i.d. random variables taking values which are equal to the roots of this linear equation is zero. Therefore, the second event happens with probability greater than 0. Since each $g_{1j}$  is a random variable drawn from a continuous distribution, det($G$)=0 happens only if the coefficients $D_{1j}=0$,  $j\in\{1,...,B\}$. Further, we can write  $\Pr (\det (G) = 0) > 0 \Rightarrow \Pr ({D_{11}} = 0) > 0$. Note that $D_{11}$  is the determinant of the matrix formed by stripping the first row and first column of $G$. Now, the same argument can be iteratively used, stripping the first row and first column at each stage, until we reach a determinant $D_{ii}$  where the next determinant along the diagonal will be unavailable. Nevertheless, we subsequently strip the first row and last column along the back-diagonal where the element is nonzero (shown as Fig. 7), until we finally reach a single element matrix containing a certain variable  $g_{Aj}$, i.e.,  $\Pr (\det (G) = 0) > 0 \Rightarrow \Pr ({g_{Aj}} = 0) > 0$. 
\begin{figure}[]
\centering
\subfigure[Stripping along the diagonal.]{
\centering
\label{Fig.sub.1}
\includegraphics[width=0.8in, angle=270]{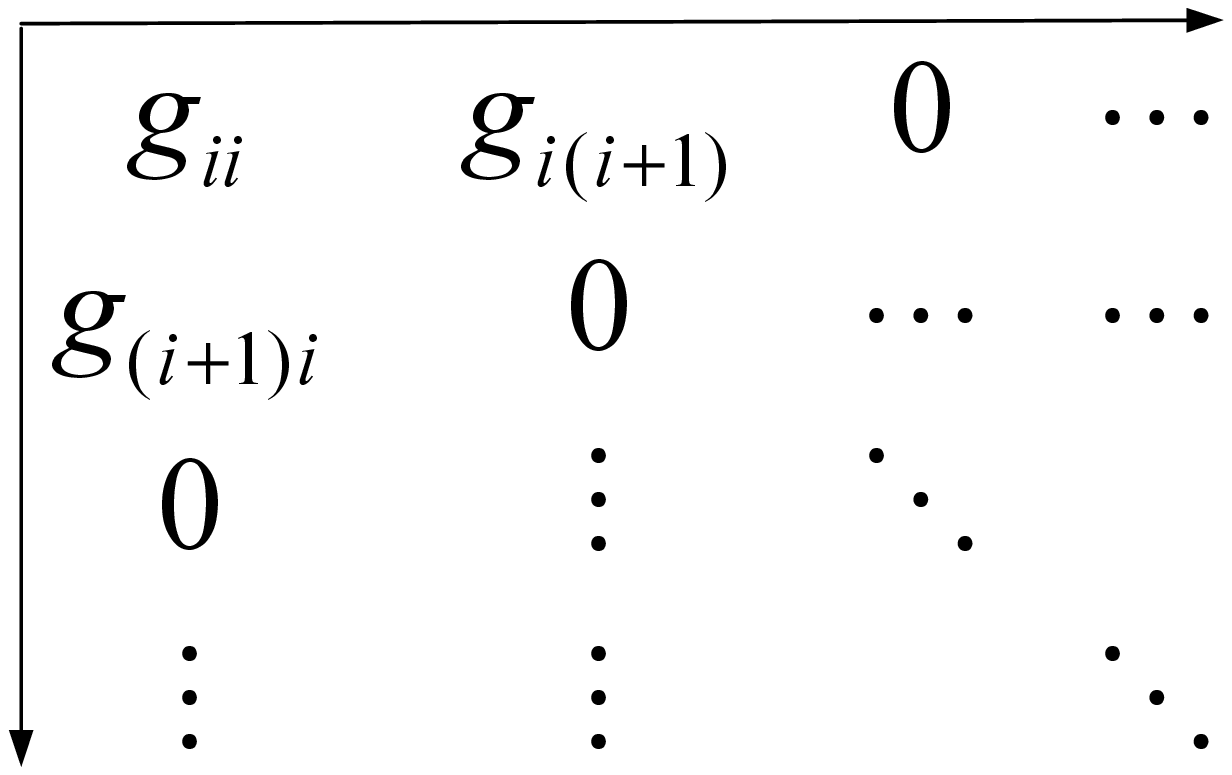}}
\hspace{0.2in}
\subfigure[Stripping along the back-diagonal.]{
\label{Fig.sub.2}
\includegraphics[width=0.8in,angle=270]{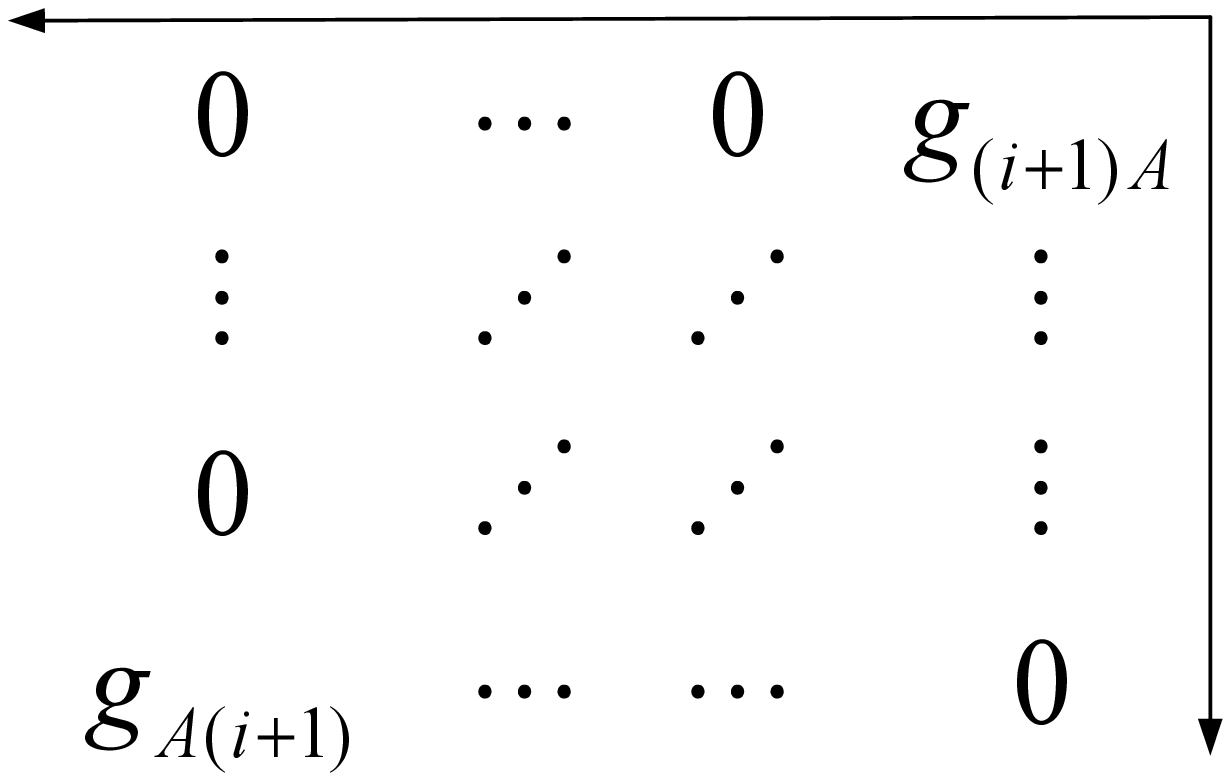}}
\caption{Case for $B$ is even.}
\label{Fig.lable}
\end{figure}

Recall that the absolute value of $g_{Aj}$  is assumed to be non-zero before. We can hence conclude that 
$\Pr ({g_{Aj}} = 0) = 0$. Thus, det($G$)  is nonzero almost surely, i.e., the matrix $G$  has a full rank of $A$ with probability 1.$\quad\quad\quad\quad\quad\quad\quad\quad\quad\quad\quad\quad\quad\quad\quad\quad\quad\quad\quad\quad\quad\quad\quad\quad\quad\quad\quad\quad\quad\quad\Box$

\textit{Cyclic Zero-Padding:} Inspired by lemma 1, if we may construct a precoding matrix in such a special form and simultaneously satisfying the condition of interference cancellation we mentioned before, each receiver will see the aligned interference shape that it previously obtained. Without loss of generality, we expound the content via (21), i.e.,  ${{\bf{H}}^{[11]}}(n){\bf{V}}_2^{[1]}(n) = {{\bf{H}}^{[11]}}({t_2})$.

Recall that the channel matrix is $B\times A$ ($B\le A$). Next, we will show that the precoding matrix may still have a full rank even if  $B<A$. By developing the expressions through the system of linear equations, we have
\begin{equation}
{{\bf{H}}^{[11]}}(n){{\bf{v}}_i} = {\bf{h}}_i^{[11]}({t_2}),
\end{equation}
where ${{\bf{v}}_i} \in {\mathbb{C}^{A \times 1}}$  and ${\bf{h}}_i^{[11]}({t_2}) \in {\mathbb{C}^{B \times 1}}$,  $i\in\{1,...,A\}$, are the $i$th column vectors of ${\bf{V}}_2^{[1]}(n)$  and  ${{\bf{H}}^{[11]}}({t_2})$, respectively. Note that in time slot $n$ we have knowledge of ${{\bf{H}}^{[11]}}({t_2})$  and ${{\bf{H}}^{[11]}}({n})$  due to the delayed and current CSIT. With the fact that the channel coefficients are i.i.d. drawn from a continuous distribution, we further have $A$ systems of linear equations and each system has $B$ linear independent equations in $A$ unknown random variables. Since  $B<A$, for the first system, let the $A-B$ unknown variables of ${{\bf{v}}_1}$  be zeroes from the bottom up in sequence and there are $B$ unknown variables left. In particular, the equivalent formula can be expressed as
\begin{equation}
{\bf{H}}_1^{[11]}(n){{\bf{\tilde v}}_1} = {\bf{h}}_1^{[11]}({t_2}),
\end{equation}
where ${\bf{H}}_1^{[11]}(n) = [{\bf{h}}_1^{[11]}(n)\,\, {\bf{h}}_2^{[11]}(n) \ldots {\bf{h}}_B^{[11]}(n)]$  is a $B\times B$  coefficient matrix containing the first $B$  columns vectors from  ${\bf{H}}^{[11]}(n)$. ${{\bf{\tilde v}}_1} = {[{v_{11}},\ldots,{v_{B1}}]^T}$  is the solution of the first system, which can be represented as  ${v_{m1}} = \frac{{{D_{m1}}}}{{\det ({\bf{H}}_1^{[11]}(n))}}$, where ${D_{m1}} = \det ([{\bf{h}}_1^{[11]}(n)\ldots{\bf{h}}_{m - 1}^{[11]}(n)\,\,{\bf{h}}_1^{[11]}({t_2})\,\,{\bf{h}}_{m + 1}^{[11]}(n)\ldots$ ${\bf{h}}_B^{[11]}(n)])$, for $m \in \{1,...,B\}$. Since the elements of $D_{m1}$  are i.i.d. drawn from a continuous distribution,  $D_{m1}$ is nonzero almost surely, i.e., the absolute value of $v_{m1}$  in ${{\bf{\tilde v}}_1}$  is nonzero almost surely. Thus, ${{\bf{v}}_1}$  can be solved in a form as  ${{\bf{v}}_1} = {[{v_{11}},\ldots,{v_{B1}},0,\ldots,0]^T}$. Furthermore, for the second system, we do the same operation as the first system but the relative positions of the zeroes. Here, starting from the last but one of  ${{\bf{v}}_2}$, let the $A-B$ unknown variables of ${{\bf{v}}_2}$  from the bottom up in sequence be zeroes, i.e., the zeroes shift one unit upwards compared with  ${{\bf{v}}_1}$. Thus, ${{\bf{v}}_2}$  can be solved in a form as ${{\bf{v}}_2} = {[{v_{12}},\ldots,{v_{(B - 1)2}},0,\ldots,0,{v_{A2}}]^T}$. We continue cyclic zero-padding until we finally reach ${{\bf{v}}_A}$  for the last system of linear equations. Therefore, we get a precoding matrix ${\bf{V}}_2^{[1]}(n)$ which is of the form as $G$  presented before. Subsequently, lemma 1 can be applied to show that such a precoding matrix has a full rank almost surely.

\bibliography{ref}

\end{document}